\documentclass[review,number,sort&compress]{elsarticle}
\usepackage{lineno}
\usepackage[ngerman,english]{babel}
\usepackage{amsmath}%
\usepackage{amsfonts}%
\usepackage{amssymb}%
\usepackage{graphicx}
\usepackage{textcomp}
\usepackage{url}
\usepackage{breakurl}
\usepackage{hyperref}
\modulolinenumbers[1]
\journal{NIM A}
\newcommand{\inch}{$^{\prime\prime}$}

\begin{document}
\begin{frontmatter}
\title{Design and realization of a sputter deposition system for the \textit{in situ}- and \textit{in operando}-use in polarized neutron reflectometry experiments}

\author[a]{Andreas Schmehl} \ead{info@Maschinen-Musik.de}
\author[a]{Thomas Mairoser}
\author[a]{Alexander Herrnberger}
\author[a]{Cyril Stephanos}
\author[a]{Stefan Meir}
\author[a]{Benjamin F\"org}
\author[b]{Birgit Wiedemann}
\author[b]{Peter B\"oni}
\author[c]{Jochen Mannhart}
\author[b]{Wolfgang Kreuzpaintner}

\address[a]{Zentrum f\"ur elektronische Korrelation und Magnetismus, Universit\"at Augsburg, Universit\"atsstr.\ 1, 86159 Augsburg, Germany}
\address[b]{Technische Universit\"at M\"unchen, Physik-Department E21, James-Franck-Str.\ 1, 85748 Garching, Germany}
\address[c]{Max-Planck-Institut f\"ur Festk\"orperforschung, Heisenbergstr.\ 1, 70569 Stuttgart, Germany}

\begin{abstract}
We report on the realization of a sputter deposition system for the \textit{in situ}- and \textit{in operando}-use in polarized neutron reflectometry experiments. Starting with the scientific requirements, which define the general design considerations, the external limitations and boundaries imposed by the available space at a neutron beamline and by the neutron and vacuum compatibility of the used materials, are assessed. The relevant aspects are then accounted for in the realization of our highly mobile deposition system, which was designed with a focus on a quick and simple installation and removability at the beamline. Apart from the general design, the in-vacuum components, the auxiliary equipment and the remote control via a computer, as well as relevant safety aspects are presented in detail. 
\end{abstract}


\end{frontmatter}


\section{Introduction}
Magnetic films and heterostructures made of them or containing them are the basic building blocks of a large number of electronic devices. Novel device approaches exploit the coupling between the electronic and magnetic degrees of freedom and thereby expand the spectrum of materials that can be used in such heterostructures \cite{10.1038/ncomms5530} to include an increasing number of elements of the periodic table \cite{doi:10.1146/annurev-matsci-070115-032057, doi:10.1021/acs.nanolett.5b02720}. The fabrication of such devices is almost exclusively based on thin film deposition techniques. As sample structure, stoichiometry, and defect population are defined during and evolve with the deposition, their precise control and optimization become increasingly important. Consequently, it is highly desirable to directly analyze the development of the physical properties of such magnetic heterostructures, e.g.\ the magnetization, during the growth process and to correlate them with the structural parameters of the sample.
The \textit{in situ} characterization of thin films by electron- and photon-based probes \cite{doi:10.1143/JJAP.29.L1199, doi:10.1063/1.1384432} as well as by scanning probe techniques \cite{doi:10.1103/PhysRevLett.64.2929, doi:10.1021/ma0119833} is common practice. For neutron based techniques, the development of \textit{in situ} setups is, however, facing complex technological challenges, as radiation protection and material activation are of high concern. 
As a spin-sensitive technique, polarized neutron reflectometry (PNR) is one of the key methods to measure the evolution of the magnitude and the orientation of the magnetic moments during growth. Hence, some approaches for \textit{in situ} PNR exist: At the J\"ulich Centre for Neutron Science (JCNS) at the Heinz-Maier-Leibnitz Zentrum (MLZ) in Garching a molecular beam epitaxy (MBE) growth chamber was built that allows the transport of the produced sample from the dedicated laboratory to the neutron reflectometer MARIA using an evacuated transport chamber. This way the sample is not exposed to ambient conditions between deposition and measurement \cite{doi:10.1063/1.4972993}. This approach allows one to specialize the deposition setup without compromising the task of thin film growth. The sample transport over long distances and through radiation protection facilities, however, slows down the measurement sequence and bears the risk of sample contamination during the transfer, transport and measurement processes. 
Another approach to exploit \textit{in situ} PNR realizes the implementation of an MBE growth chamber within a compact, light-weight, and portable design directly at the sample position of a neutron reflectometer \cite{doi:10.1063/1.3169506}. In this design, the thin film sample is moved by a carrier over short distances from a growth position to the PNR analysis position inside the very same vacuum chamber. The advantages over the aforementioned approach is the possibility of a much faster sequence of deposition and analysis. This reduces the risk of contamination of the sample and influences on the measurements originating from differing physical environments at the coating and measurement positions. Nevertheless, a PNR measurement during the film deposition (i.e.\ \textit{in operando}) cannot be realized using this design as the sample is not kept aligned in the neutron beam for the coating process.
Ultimately an optimized \textit{in situ} PNR experiment would allow measurements during deposition or in short breaks in between deposition runs to be made without any change of sample position and beam alignment. In addition to a deposition system that is suited for this approach, a brilliant neutron beam is required at the sample position that present day neutron sources cannot provide. However, the massively increased brilliance of the future European Spallation Source ESS in combination with improved neutron optical concepts \cite{doi:10.1051/epjap/2012110295, doi:10.1016/j.nima.2016.03.007} will soon allow for \textit{quasi in operando} experiments, where only an interruption in sample growth of few seconds is required for the PNR analysis.
Therefore, we decided to develop a thin film deposition system which allows the sample to remain aligned in the neutron beam for alternatingly carried out deposition and measurement processes. The system closes the gap to fully \textit{in operando} experiments and opens up new technological capabilities for today’s \textit{in situ} PNR. This system allows for the first time to explore important aspects of material activation, size limitations, vibrational influences, and geometrical restrictions.
This work describes the design of a thin film deposition system for \textit{in situ} and \textit{in operando} PNR experiments. First scientific result obtained with this coating setup have recently been published elsewhere \cite{doi:10.1103/PhysRevApplied.7.054004}.

\section{System Specifications}
\subsection{General Design Considerations}

Because of spatial limitations and flexibility considerations, we chose magnetron sputtering as the deposition technique. Besides the necessities that come with the chosen deposition method, a large spectrum of general and experiment-specific limitations have to be taken into account when designing a deposition system aimed for the use in a neutron beamline.

\subsection{The REFSANS Beamline}
\label{REFSANS}
\begin{figure}
	\includegraphics[width=\columnwidth]{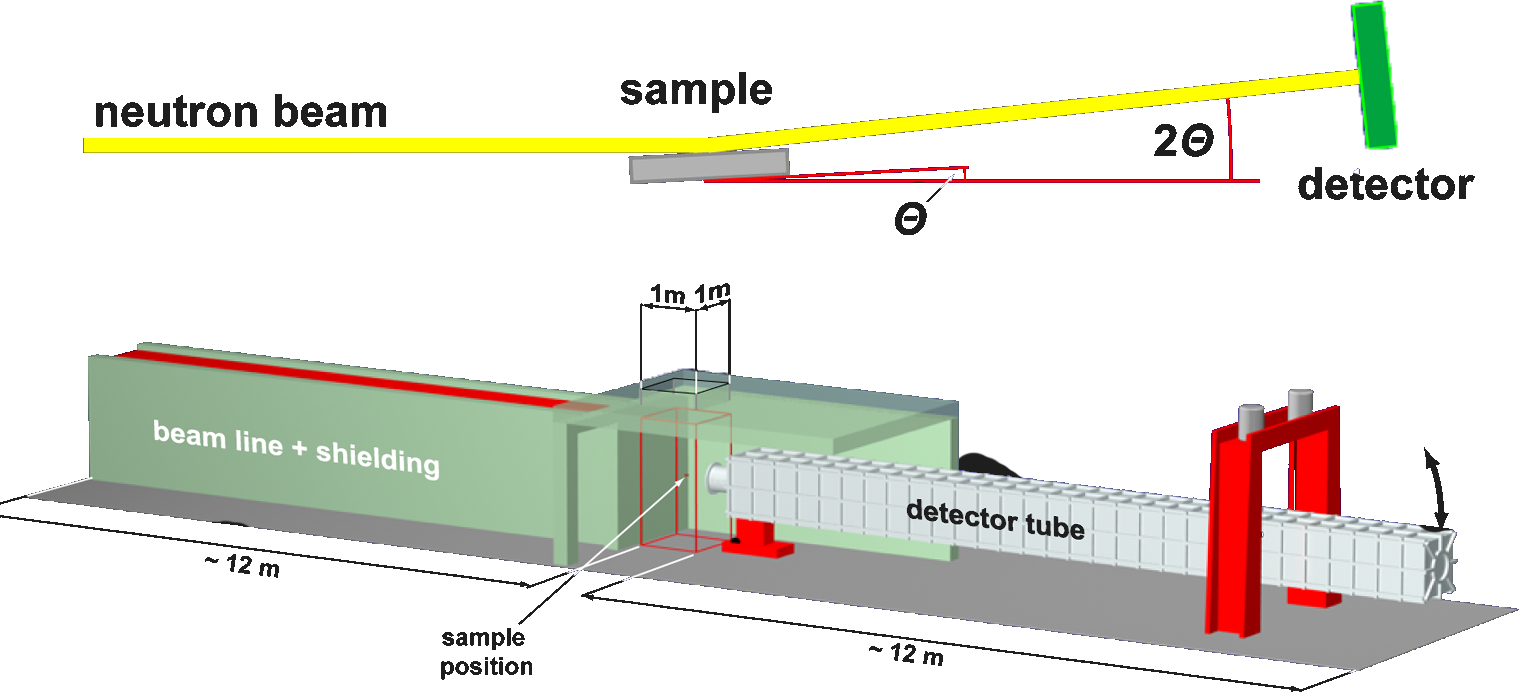}
	\caption{Schematic of the REFSANS beamline at the FRM\,II with $\theta - 2\theta$ scattering geometry, where the incidence angle of the neutrons is defined by the inclination of the sample ($\theta$) against the horizontal. The geometry of the PNR experiment and the spatial limitations of the experiment chamber were leading in the design considerations of the deposition system.}
	\label{Fig1}
\end{figure}

The deposition system was originally designed to be used with the REFSANS beamline operated by Helmholtz-Zentrum Geesthacht at the MLZ \cite{Kampmann20061161, Kampmann2004E763, Moulin20151}. The beamline's geometry and its spatial restrictions were, therefore, the driving factors behind the design. As REFSANS serves in user operation, the beamline could not exclusively be reserved for permanent installation of the deposition system, which was, therefore, designed mobile and with a focus on a quick and simple installation and removability. This lead to a system design that has a very low footprint and very flexible adjustment, both features that also enable its operation at other beamlines. 
Mainly designed to allow the analysis of soft and liquid materials, but also highly suitable for the investigation of solid state samples and thin films, REFSANS utilizes a horizontal sample geometry (Fig.\ \ref{Fig1}) with vertical scattering plane (defined by the incident beam and the detector arm). For solid samples, neutron incident angles of up to more than 10$^{\circ}$ are possible, with a maximum irradiated sample area of approximately $ 65 \times 65$\,mm$^2$.
To utilize the full angular range of the reflectometer, any sample environment (such as a thin film deposition system), which due to its size replaces the available standard goniometer, has to offer its own accurate and reproducible height adjustability in the range of about 200\,mm with an accuracy of better than 0.1\,mm. It also must offer the possibility to correct for unwanted tilt angles of the samples relative to the scattering plane (usually referred to as $\chi$- or ``roll''-axis) with an angular accuracy of better than 0.1$^{\circ}$. This is usually achieved by installing the samples on a purpose designed eucentric, two-axes goniometer with an additional z-axis for precise height adjustments.

The actual experiments on REFSANS take place in an access limited experiment area of about $3 \times 4$\,m$^2$ lateral size, enclosed by several cm thick steel walls. It is accessible through two doors at floor level and through a top hatch, which is intended for the installation of larger experiment equipment using the neutron guide hall's gantry crane. The dimensions of the top hatch are $100 \times 95$\,cm$^2$ and define the strict size limit for the equipment to be installed through it. Additional space is available in the experimental area on a small, about 1\,m wide hallway next to the actual beam path. 
Due to neighboring instruments, direct access to the experiment area via the ground level is also limited. The equipment, that is not loaded through the top hatch, either has to be sufficiently small or dismountable. Alternatively, equipment can be positioned next to the experiment area or on top of the neutron optics casemate. 
Because neutrons carry a magnetic moment, magnetic stray fields can have a detrimental effect on diffraction and reflectometry experiments. This situation becomes especially relevant, if the respective magnetic fields also show fast (non-adiabatic) variations. Therefore, one has to ensure that at the sample position, the measurement equipment generates no relevant magnetic fields. Unavoidable fields have to be shielded. In addition, well defined static magnetic fields at the sample position can be used to compensate or dominate unwanted stray fields of the equipment. Both approaches, field minimization and defined static fields are included in the design. 

\subsection{Material Limitations}
The high vacuum environment of a sputter coating system, as well as the neutron radiation impose strict limitations on the choice of materials that can be used for a deposition system. One of the major concerns is to avoid radioactive activation of the materials exposed to the neutron beam.
This can be achieved by choosing materials with low neutron capture cross sections, by using materials that do not become radioactive after neutron capture, or by shielding parts of the equipment against neutrons. In addition to these material limitations, the chosen materials have to be compatible with the ultra-clean, high vacuum, and high temperature environment of a modern sputter deposition system. To avoid outgassing and to ensure short pump-down times, low base pressures, and negligible sample contamination, dense materials with low porosity have to be used. Furthermore, the employed materials must be able to sustain high temperatures in the vicinity of the substrate heater. The same holds for the materials that are exposed to either inert gas or oxidizing sputter plasma. To assess the usability of materials for a neutron radiation environment one has to compare their neutron capture cross sections and the decay times of the resulting radioactive isotopes. A good quantity to compare the possible radioactive activation of materials is the storage time $t_{\text{store}}$ which is needed for a sample of 5\,cm$^3$ volume to decay to an activity of 74\,Bq/g after exposure to a neutron yield of $10^7$\,n\,cm$^{-2}$\,s$^{-1}$\,\cite{NeutronDataBooklet}. At 74\,Bq/g or less a sample is declared as non-radioactive. The values of the storage times, capture cross sections, and decay times are tabulated. As the deposition system has to be mobile and manually accessible at all times, the materials restrictions play a dominant role in its design. Following this assessment, one can categorize materials into three groups: (i) Materials that must not be exposed to neutron radiation, (ii) Materials that can be exposed to low doses of neutron radiation and (iii) Materials that do not become radioactive under neutron radiation. Their specific characteristics are outlined in the supplementary material provided.

\subsection{Deposition System Specifications}
Besides the requirements imposed by the beam line geometry and the neutron radiation environment, the design of the deposition system is also governed by the need to realize high-quality epitaxial thin films and heterostructures out of a very broad spectrum of materials. The desired applications and requirements of the deposition system were specified as follows: 

\subsubsection{Sample Heating}
As many scientifically interesting materials can only be grown epitaxially at high temperatures, a substrate heater is required. To maximize the sample surface for the reflectometry measurements, and thus to reduce data acquisition times, the substrate heater was specified to hold samples up to 72\,mm in diameter. This allows for a maximum square surface illuminated by the neutron beam of $50 \times 50$\,mm$^2$. To give access to a wide thermodynamic range, the heater was specified to operate up to a maximum temperature of 1000$^{\circ}$\,C in inert gas and up to 800$^{\circ}$\,C in oxidizing atmospheres.

\subsubsection{Sputter Facility Requirements}
The demand to deposit insulators, semiconductors and metallic conductors necessitates the use of both direct current (DC) and radio frequency (RF) sputtering. The materials of interest encompass metals, alloys, metal oxides and possibly nitrides. Therefore, the system was specified to use an inert argon plasma for non-reactive sputter deposition, as well as combinations of argon/oxygen or argon/nitrogen plasmas for reactive sputter processes. As many of the most interesting physical effects occur at interfaces between or within heterostructures of different materials, the system was specified to allow up to three different materials from independent sputter sources to be deposited \textit{in situ} without breaking the vacuum. 

\subsubsection{Sample Positioning}
To be able to deposit films under optimized sputter conditions, the distance between sputter source and substrate was made adjustable. For two inch-diameter sources, the desirable distance range is about $40 - 90$\,mm. This requires a vertical adjustability (z-axis) of $\pm 25$\,mm relative to a central position. To furthermore ensure optimized conditions for the neutron reflectometry measurements, additionally a lateral adjustment along the beam axis (x-axis) of $\pm 25$\,mm is also mandatory. The need to ensure a perfect planar alignment of the sample surface with respect to the beam geometry necessitates the implementation of a high-resolution sample rotation (d-axis) perpendicular to the scattering plane. 

\subsubsection{Remote Controllability}
The deposition system is designed to allow \textit{in situ} PNR measurements to be performed in a sequences of film growth and neutron exposure steps. Because of radiation safety, the system is not directly accessible during the experiments, such that all relevant functions of the system have to be controlled remotely. This includes the sample positioning and heating, the sputter gun control, the gas handling, the pump control, rate-, pressure-, and temperature-monitoring, as well as the control of the vacuum system itself. 

\subsubsection{Mobility Requirements}
Due to the boundary conditions outlined in section \ref{REFSANS}, for an effective use of allocated beam time, and to ensure a quick and easy set-up and removal, on-site assembly requirements were minimized by splitting the system into pre-assembled, easy to handle and to connect modules that can be positioned using the available gantry crane.

\section{Realization}

\subsection{Geometry of the Vacuum System}
The general geometry of the system is determined by the need to have a fixed sample position. This demands a system layout, in which the three sputter sources can be changed with respect to the sample, while the sample stays aligned in the neutron beam. MeiVac 2\,\inch\ MAK magnetron sputter sources were chosen as they allow both RF and DC operation under a wide spectrum of sputtering conditions. Their permanent magnets are made of a neodymium/iron/boron alloy with a very short storage time, should a source accidentally be exposed to the neutron beam. With the exception of small amounts of silver solder, the sources are made from materials that are compatible with the neutron environment within the deposition chamber. The sputtering targets are attached magnetically to the sources, which allows for easy and quick target exchange, even in the confined space of the experiment hutch.

\begin{figure}
	\includegraphics[width=\columnwidth]{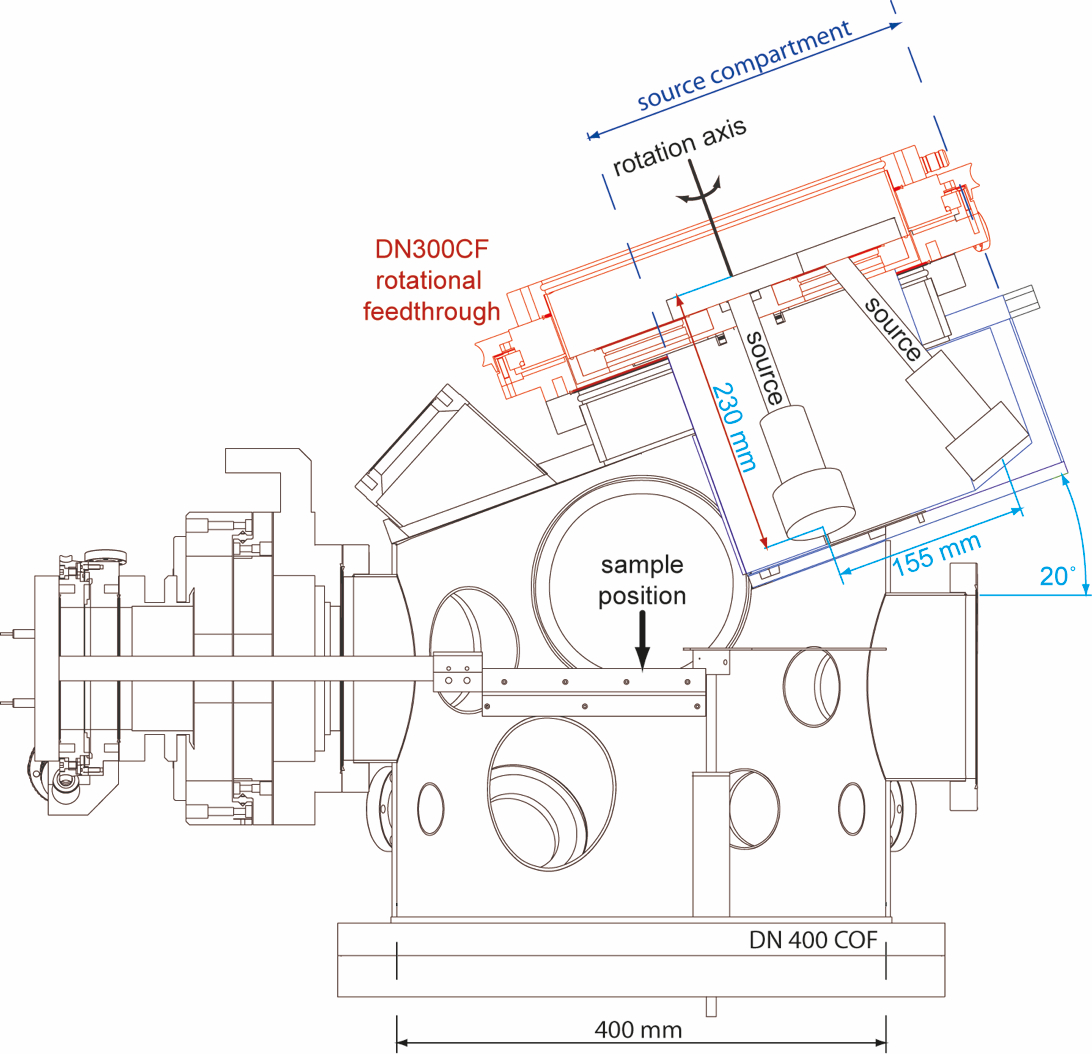}
	\caption{Sketch of the sputtering system showing the configuration of the sputter sources. The sample position stays fixed and the sources are positioned above the sample by rotating them via the feedthrough. To maximize the source spacing, they are designed to be 230\,mm long and to be tilted by 20$^{\circ}$ with respect to the sample plane. Thereby they rotate on a 155\,mm radius circle. Depicted is the blank position for reflectometry experiments.}
	\label{Fig2}
\end{figure}

To position the sources with respect to the sample, a revolver-like design was chosen with the sources mounted on a rotational vacuum feedthrough above the sample (Fig.\ \ref{Fig2}). To minimize any influence of a magnetic stray field of the magnetron sputtering sources at the sample position during a PNR measurement, a fourth, intentionally left blank, virtual source position was implemented. It is moved above the sample for the neutron beam exposure, such that all sputtering sources are positioned away from the sample. Because of the spatial limitations, the footprint of the rotation mechanics was minimized. A custom designed 300\,mm diameter feedthrough was chosen, just large enough to host four DN63CF flanges for the installation of the sources. It was fabricated to standard DN-CF specifications\footnote{VAb Vakuum-Anlagenbau GmbH, Marie-Curie-Str. 8, Postfach 1146, 25337 Elmshorn, Germany} and has a two-stage differentially pumped sealing system. Rotation is realized by a stepper motor via a worm gear. To avoid cross contamination, the sources have a distance of 230\,mm between source face and base flange. Additionally, the source are tilted outwards by 20$^{\circ}$ with respect to the feedthrough (Fig.\ \ref{Fig2}). Hence, whereas the source base flanges are arranged on the feedthrough on a 67.5\,mm radius circle, the elongation and the tilting put the source faces on a 155\,mm radius circle, maximizing the distances between them. As for the deposition process the source face (i.e.\ the target face) has to be parallel to the sample surface, the complete source manipulation assembly was again tilted by 20$^{\circ}$ with respect to the horizontal plane. Thereby, the source positioned above the sample is in vertical alignment. 

\begin{figure}
	\includegraphics[width=\columnwidth]{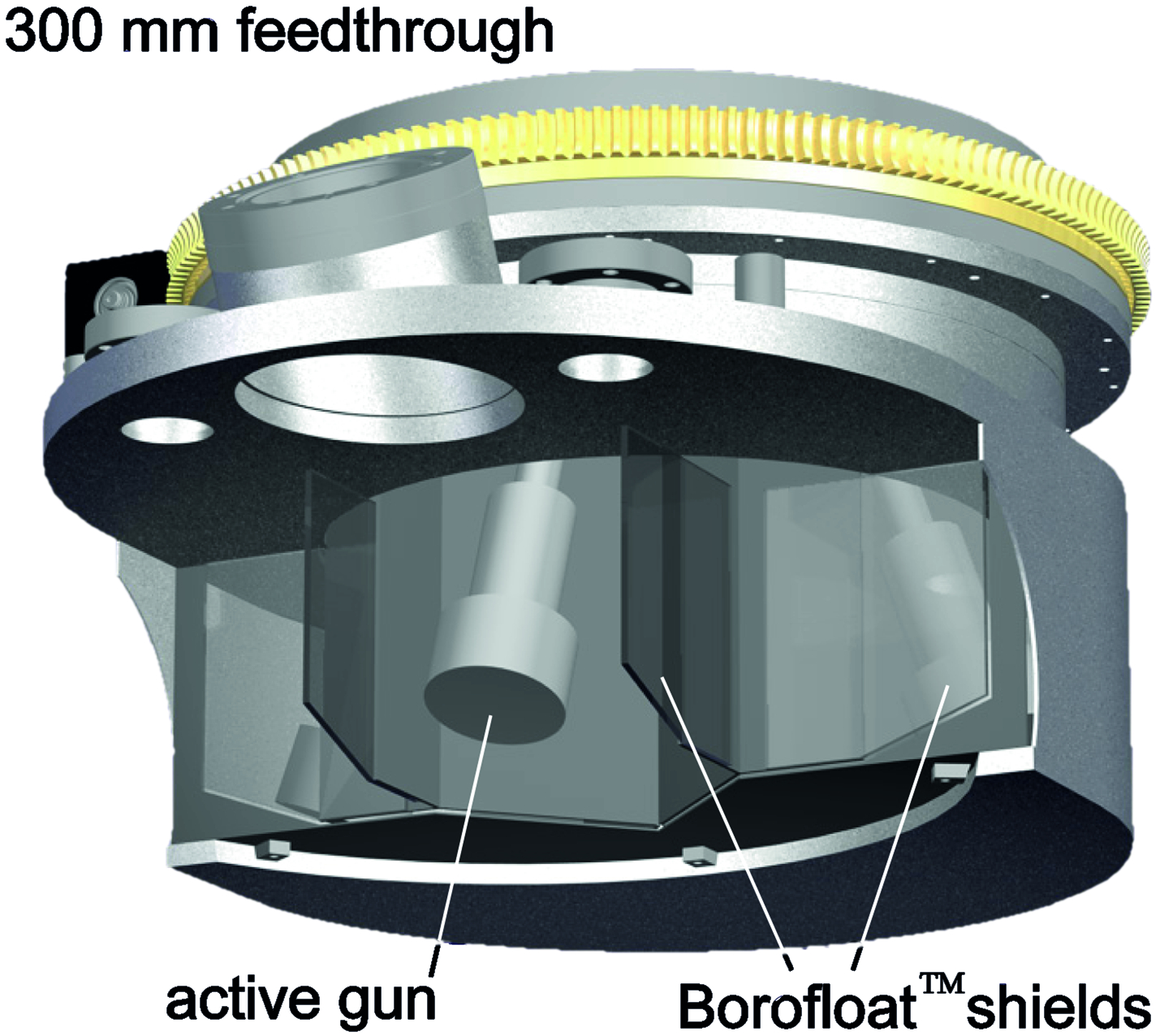}
	\caption{CAD drawing of the source compartment of the deposition system, with the 300\,mm rotational vacuum feedthrough on top. The active, central source is protected by a pair of shields from neutron radiation passing through the system from left to right. The remaining two sources (left and right) are stored in the compartment section, protected from cross contamination by their individual pairs of shields. The neutron shielding is achieved by the use of Borofloat\texttrademark.}
	\label{Fig3}
\end{figure}

To further reduce the risk of cross contamination between the individual sputter sources, they are separated from each other by 6\,mm thick Boron silicate glass (Borofloat\texttrademark) shields, arranged in a star-like pattern around the rotation axis of the rotational feedthrough. The shields create sections which the sources are located in (Fig.\ \ref{Fig3}) and close off the separate source compartment against the main chamber. This design improves the shielding of the sputter sources as only the source in use protrudes into the sample section of the chamber (Fig.\ \ref{Fig2}). To prevent unwanted radioactive activation of the sputtering sources, the boron silicate glass is also thick enough to protect even against a direct neutron beam exposure. Boron silicate glass was chosen as it can withstand the sputtering plasma and the heat radiation of the sample heater, while staying inert to oxidizing atmospheres and plasmas.

\begin{figure}
	\includegraphics[width=\columnwidth]{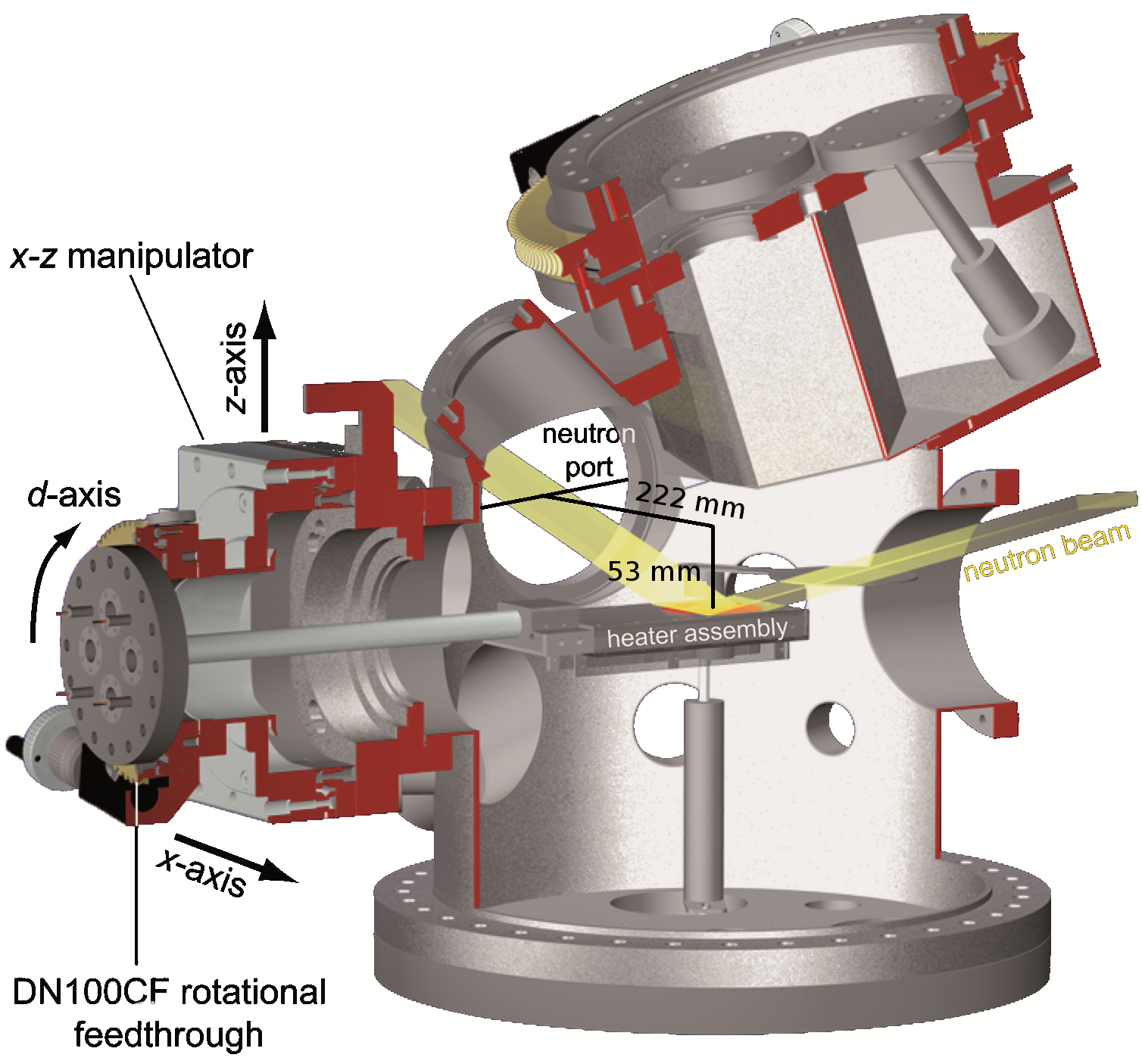}
	\caption{Virtual cut through the deposition system (radius: 222\,mm) in its reflectometry measurement configuration. The neutron beam enters through the left DN160CF port and exits through the right one (not shown) in the range of $−2^{\circ}$ to $+25^{\circ}$ with beam cross sections of $50 \times 10$\,mm$^2$. The sample, mounted on the heater assembly, is positioned 53\,mm below the flange axis within the beam. The sample tilt is adjusted with respect to the incident beam via a DN100CF rotational feedthrough (d-axis).}
	\label{Fig5} 
\end{figure}

The deposition system combines two main elements, the upper section, housing the described 20$^{\circ}$ tilted source compartment, and a lower, cylindrical section, made of a 400\,mm diameter stainless steel tube (Fig.\ \ref{Fig2}) which ends in a 400\,mm COF wire seal flange, that also forms the base of the vacuum chamber. This lower flange is equipped with a set of DN40CF and DN63CF flanges to attach auxiliary equipment like bypass and overpressure valves. 
The cylinder section of the vacuum chamber opens into four DN160CF flanges, spaced 90$^{\circ}$ horizontally around the cylinder axis with respect to the neighboring flange. The left and right flanges are aligned perpendicular to the scattering plane and act as the ports for the incoming and reflected neutron beams. They are positioned such that they allow the specified beam geometry of PNR experiments to be fulfilled: incident and reflected beam angles from $-2^{\circ}$ to $+25^{\circ}$ and a beam of 50\,mm width and 10\,mm height. In detail, this yields a flange-to-flange distance of 444\,mm and a sample position 53\,mm below the flange axis (Fig.\ \ref{Fig5}). The flanges are closed with in-house fabricated neutron windows composed of 10\,mm thick boron free glass with a neutron transparency of about 90\,\%. As sputter coating of these windows might over time affect their neutron transparency, they are designed for quick replacement.

Parallel to the scattering plane and half way between the two neutron window flanges, a central DN160CF flange is located which supports the heater manipulators and the heater assembly. Its position with respect to the z-axis was determined by the requirement to adjust the sample-to-sputter source distance between $40 - 90$\,mm. The relative positions of the neutron port flanges, the heater support flange and the source assembly were then chosen such that for the standard sample-to-sputter source distance of 50\,mm the sample also is positioned correctly for the reflectometry measurements. Attached to this central flange is a two axis UHV linear motion feedthrough for the positioning of the sample along the x- and z-axes within the specified $\pm 25 \times \pm25$\,mm$^2$ window. Both linear axis of the manipulator are driven by high accuracy stepper motors. At x = 0\,mm the sample is located exactly below the source positions in the center of the vacuum chamber. 

Attached to the x-z linear motion feedthrough is a 100\,mm diameter UHV DN100CF rotational feedthrough that supports the heater assembly. For a highly accurate rotation of the heater around the d-axis, the feedthrough is actuated by a stepper motor via a worm gear (Fig.\ \ref{Fig5}). This rotation is employed to align the sample surface with respect to the beam geometry and to rotate the complete assembly by 180$^{\circ}$ for the rate monitoring (see Fig.\ \ref{Fig10}). 
The system is further equipped with a multitude of additional flanges of various sizes and orientations to support auxiliary equipment (e.g.\ pressure gauges, pump ports, feedthroughs, etc.). They are also used as mounting positions for various optional equipment such as beam apertures and magnetic field coils.

\subsection{System Mounting and Height Adjustment}
The sample can only be tilted around a single axis (d-axis). Hence, the deposition system has to be adjustable to ensure the alignment of the sample with respect to the scattering plane.
This was realized by mounting the vacuum system on a rigid steel support frame which is gimbal-mounted by four 16\,mm diameter spherical swivel heads. The latter are attached to length adjustable M16 thread rods. Changing of the mounting lengths of the swivel heads results in a tilting of the deposition system around two perpendicular axes, defined by the two pairs of swivel heads. The optimal position is then secured by countering the threads with M16 nuts. 
To operate the reflectometer over the complete range of possible neutron incident angles, the vacuum chamber also has to be adjustable in height (z-axis) within a 20\,cm range and with about 0.1\,mm accuracy. A special chamber manipulator was designed, that is located between the base plate of the system (typically screwed to the beamline mounting rails) and the gimbal-mount of the vacuum chamber. As this manipulator has to support the $\geq 400$\,kg weight of the deposition system and has to be compatible to the accuracy and mobility requirements of the system, a very robust and rigid design has been developed (Fig.\ \ref{Fig6}):

It is based on a standard manual forklift with 1000\,kg lifting capacity. This forklift was cut down to respect the size limitations of REFSANS, and was then welded onto the 1\,cm thick steel base plate of the system. The pneumatic lift actuator was replaced by a servo motor actuated lift drive, that operates via a jack shaft. It has a lifting capacity of 1000\,kg and achieves the specified positioning accuracy. The lift drive is screwed to the system’s base plate, and its jack shaft is attached to the mobile fork. The steel struts, which support the swivel heads of the chamber mounting, are welded to the shortened forks of the forklift. The chamber weight pretensions the lifting system, eliminating all possible backlash that could originate from the forklift design. 
The outer dimensions of the complete assembly that has to be installed through the top hatch of the REFSANS experiment are defined by the system mounting (with lift drive) and the sputter deposition system (Fig.\ \ref{Fig6}). The latter is topped off by the RF sputtering matchbox, which automatically controls the impedance mismatch between RF power supply and sputter plasma. As the matchbox has to be installed as close to the sputter sources as possible, it was made part of the system assembly. With a width of 86\,cm and a length of 96\,cm the complete deposition system fits through the experiment top hatch with  $\sim 2$\,cm gaps to either side. With 148.5\,cm it is also sufficiently small in absolute height to be handled via the neutron guide hall's gantry crane. 

\subsection{In-Vacuum Components}
\subsubsection{Heater and rate monitor}
The central part of the sputter deposition system is the sample heater and the rate monitoring sensor, which are combined into one assembly. As it is part of the UHV vacuum system, has to operate at high temperatures, under direct neutron radiation, and oxidizing atmospheres, new material solutions and layouts were developed to meet these challenging requirements.

\paragraph{Heating Element}
The heater is designed with maximum flexibility with respect to substrate types, shapes and mounting techniques. Therefore, a radiative layout was chosen, in which the thermal radiation of a resistive heating element is used to either directly heat the substrate or a supporting puck, on which the substrate is located. Because of the high transparency of most materials to neutrons, a major fraction of the incident beam will not be scattered by the sample, but will irradiate the sample puck and the heating element below. To avoid neutrons being scattered in the sample mount and/or heating element, and to prevent radioactive activation, both heater parts were fabricated from neutron absorbing materials. Because of the excellent temperature stability, vacuum compatibility and thermal conductivity, B$_4$C was chosen as sample puck material. The heating element is based on a boron nitride coated pyrolytic graphite resistor element (Boralectric$^{\text{\textregistered}}$), that was produced to our specifications \cite{tectra,Momentiveperformanceproducts} (Fig.\ \ref{Fig8}) to ensure the best possible temperature uniformity over the heated surface and to minimize the magnetic field generated by the heating current. The latter is achieved by realizing the resistive tracks in a counter-wound configuration, such that the magnetic fields generated by neighbouring tracks cancel each other out. 
For a high temperature uniformity and constant energy dissipation all resistive tracks have to be of the same width (assuming a constant track thickness). To minimize the required power input, additionally a maximum surface coverage of the heating element by the tracks is desired. These requirements could be met by a track geometry that is based on two interlaced spirals with opposite rotational direction. The design and the resulting current directions are depicted in figure 8. The twin-spiral design has a constant track width in the hot region of the heating element and thus constant current density and heat dissipation. Furthermore almost 100\,\% surface coverage of the hot region below the sample puck could be realized. 
Because of a pBN capping, most of the heating element’s graphite resistor is well protected from oxidation at high temperatures.

Nevertheless, the exposed graphite contacts for the electric connection can be attacked by oxygen, limiting the life time of the heating element. To minimize this effect, the heat load on the contacts was reduced by positioning the contacts far away from the hot region, resulting in an L-shape of the heating element. The conducting paths leading to the contacts are also substantially wider than in the hot region, resulting in a reduced energy dissipation per surface area. Additional protection against oxygen is provided by graphite washers, which protect the exposed contact area.

\begin{figure}
	\includegraphics[width=\columnwidth]{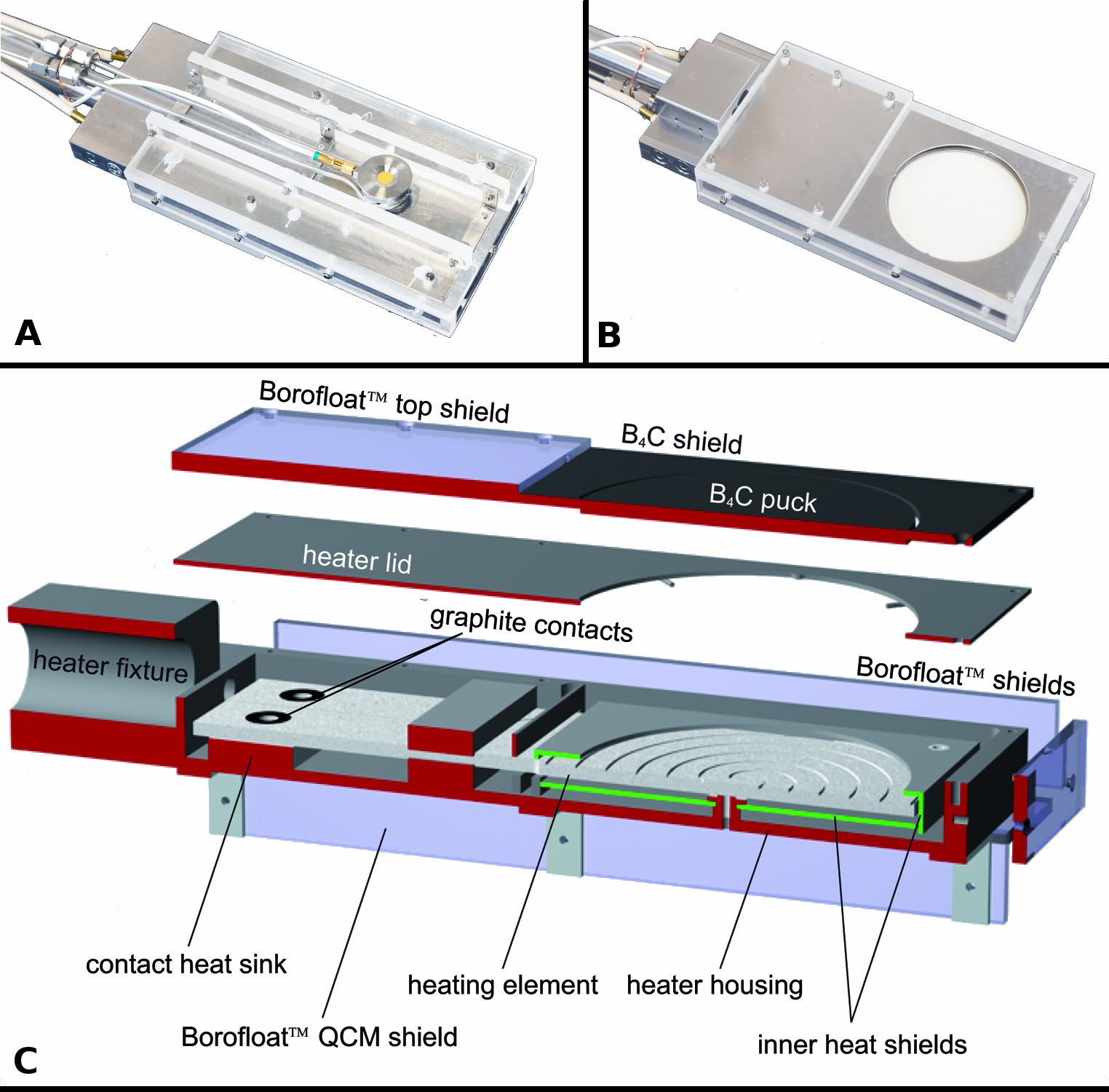}
	\caption{\textbf{A}: Photograph of the backside of the heater with the rate monitor sensor head in the central position. Two steel tubes provide cooling water for the sensor. \textbf{B}: Photograph of the sample side of the heater assembly with the white heating element showing through the top heat shield (B$_4$C neutron shields not attached). For neutron radiation protection, the whole assembly is covered in Borofloat\texttrademark\ glass. \textbf{C}: Exploded and cut view of the heater assembly (CAD rendering). The hot part of the heating element (right half of assembly) is enclosed in inner and outer radiation shields. The graphite contacts of the heating element are located in the cold part of the heater and are further cooled through a thermally coupled heat sink. Radiation shielding for the heater and the rate sensor (QCM) are provided by Borofloat\texttrademark\ glass and by B$_4$C in the hot areas.}
	\label{Fig10}
\end{figure}

\begin{figure}
	\includegraphics[width=\columnwidth]{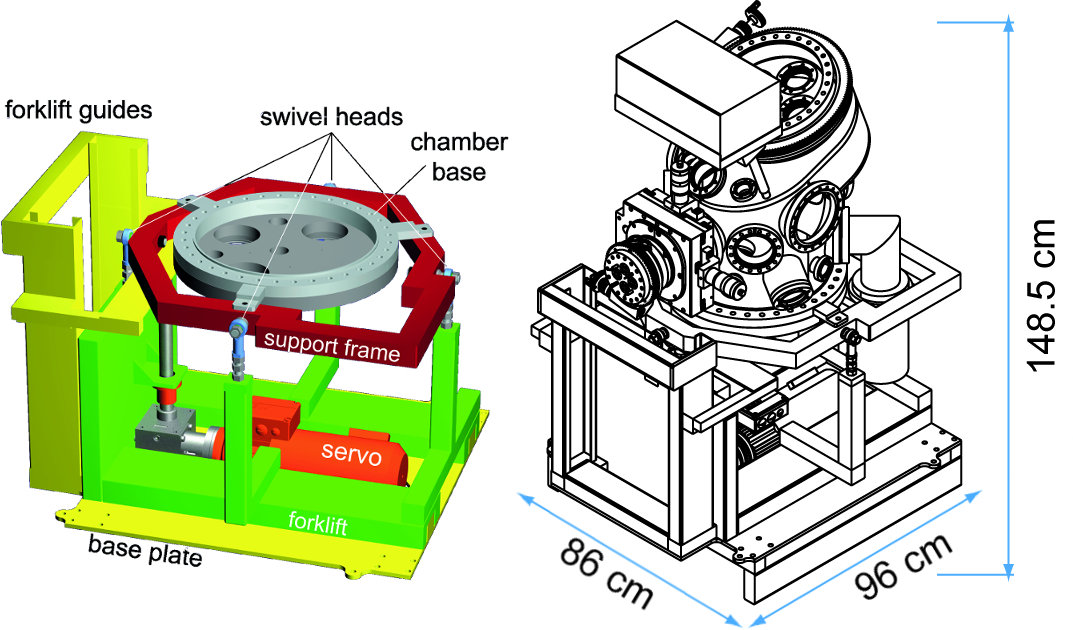}
	\caption{\textbf{Left}: CAD rendering of the chamber lift system which is based on a standard laboratory forklift. The forklift was cut to the desired dimensions, welded to a steel base plate and equipped with a servo lift drive. The gimbal–mounted support frame allows the leveling of the deposition chamber along two axes. \textbf{Right}: 3-dimensional view of the complete deposition setup consisting of the vacuum system and the forklift mounting. The footprint of $86 \times 96$\,cm$^2$ falls within the $100 \times 95$\,cm$^2$ limit of the top hatch of the REFSANS beamline at MLZ. Its comparatively small height of 148.5\,cm allows for its manipulation with the experiment hall gantry crane.}
	\label{Fig6}
\end{figure}

\begin{figure}
	\includegraphics[width=\columnwidth]{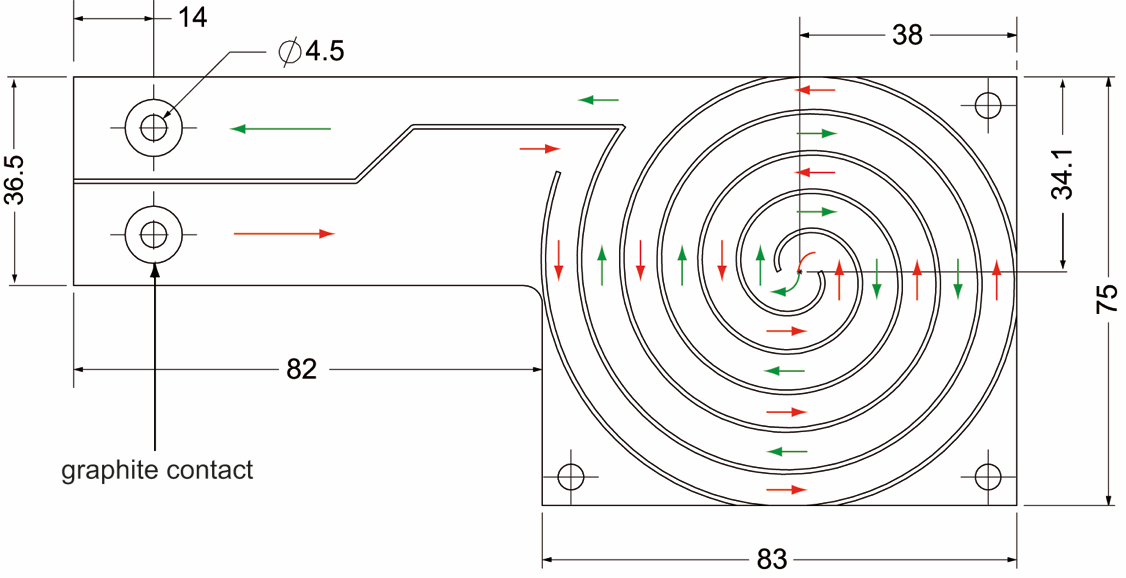}
	\caption{Top view of the boron nitride based heating element. The resistive tracks are configured in a counter-wound spiral configuration that minimizes magnetic field generation by the heating current. The in and out current paths are depicted by red and green arrows, respectively. The constant track width provides a constant energy dissipation and thereby a homogenous temperature across the surface area. To provide cooling, the graphite contacts are spatially separated from the hot zone. Distances are given in mm.}
	\label{Fig8}
\end{figure}

\paragraph{Heater Assembly}
Due to spatial restrictions and to minimize the area exposed to neutrons, special emphasis on a low-profile design of the heater was taken. To minimize the required input power and the energy loss through radiative heat dissipation, the heating element is surrounded by a sequence of two stainless steel radiation shields. 
The inner set of shields completely covers the hot area of the heating element, and only allows heat radiation to escape through a circular opening in the top, which has a slightly larger diameter than the sample puck. The pBN heating element is suspended in between (Fig.\ \ref{Fig10}). Thermal insulation between the heating element and the shields is provided by hollow Macor\texttrademark\ spacers. The second set of shields is formed by the heater housing, the heater cap and a thin steel heat-barrier within the heater. The latter divides the interior of the heater into a hot and a cold area, where the contacts of the heating element are located. To aid in preventing the oxidation of the graphite contacts, the contact area of the heating element is thermally coupled to the significantly colder heater housing via a steel block. This acts as a heat sink (Fig.\ \ref{Fig10}). Acting also as the sample holder, the heater parts are protected from radioactive activation by neutron absorbing shields. In the colder areas of the heater ($T < 500^{\circ}$\,C), this shielding is provided by 6\,mm thick Borofloat\texttrademark\ glass sheets, which are screwed to the heater housing and completely cover the assembly. In the hoter areas of the heater, near the B$_4$C puck, the shielding is provided by 2\,mm thick laser cut B$_4$C (Fig.\ \ref{Fig10}). To measure the deposition rates of the sputtered materials, a quartz crystal microbalance (QCM) \cite{QCM} is used. To get most accurate readings, 
the QCM sensor head is installed in the sample axis on the backside of the heater (Fig.\ \ref{Fig10}, top). By rotating the heater by 180$^{\circ}$ via the d-manipulator, and by adjusting the z-position of the heater assembly, the QCM can be moved into the sample position. This allows for the direct measurement of deposition rates without the need for geometric conversion factors. The QCM sensor and its electric and cooling water feeds are protected against neutron radiation using Borofloat\texttrademark\ shields (Fig.\ \ref{Fig10}). The complete heater assembly is mounted on a 15\,mm diameter steel axis, which together with electrical and water feedthroughs is centered on the DN100CF flange of the d-axis rotational feedthrough. 

During operation, the heater shows a homogenous temperature distribution across the heating element and the B$_4$C sample puck. 
A moderate input power of about 600\,W is needed to reach a sample puck surface temperature (measured by pyrometry) of 800$^{\circ}$\,C. Permanent operation at 800$^{\circ}$\,C leads to a saturation temperature of the vacuum system outer walls in direct vicinity of the heated area of about 60$^{\circ}$\,C, well within the limits of the equipment. 

\subsubsection{Guide fields and in-vacuum beam aperture}
\begin{figure}
	\includegraphics[width=\columnwidth]{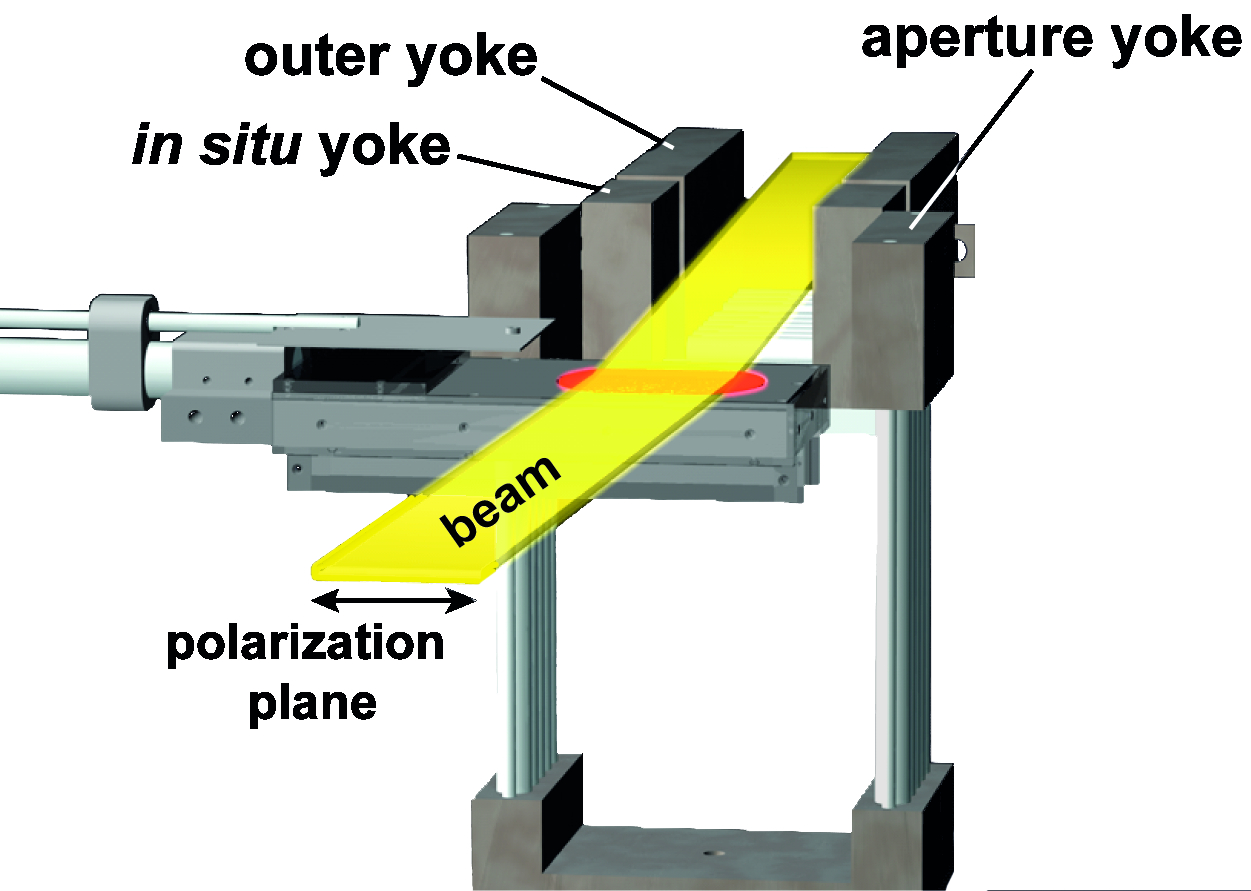}
	\caption{CAD rendering of the magnetic yoke setup providing the guide field for the polarized neutron beam. The outer yoke is placed between the exit window of the neutron beamline and the entry window of the deposition chamber. The inner yoke and the aperture yoke bridge the distance between chamber entry window and sample holder. The wider design of the aperture yoke allows the installation of the \textit{in situ} neutron beam aperture (not shown).}
	\label{Fig12}
\end{figure}

To rule out negative influences on the neutron beam polarization originating from stray magnetic fields of the auxiliary equipment, a static horizontal magnetic guide field of about 10\,G, aligned parallel to the y-axis of the chamber between the beam exit window of the beamline to the sample position was implemented. The field is generated by strong NdFeB permanent magnets and is aligned using iron yokes on the outside and the vacuum side of the chamber (Fig.\ \ref{Fig12}). The exact magnetic field configuration has been designed using finite element (FE) Maxwell equation simulations and was verified using Hall sensor measurements (Fig.\ \ref{Fig15}).

\begin{figure}
	\includegraphics[width=\columnwidth]{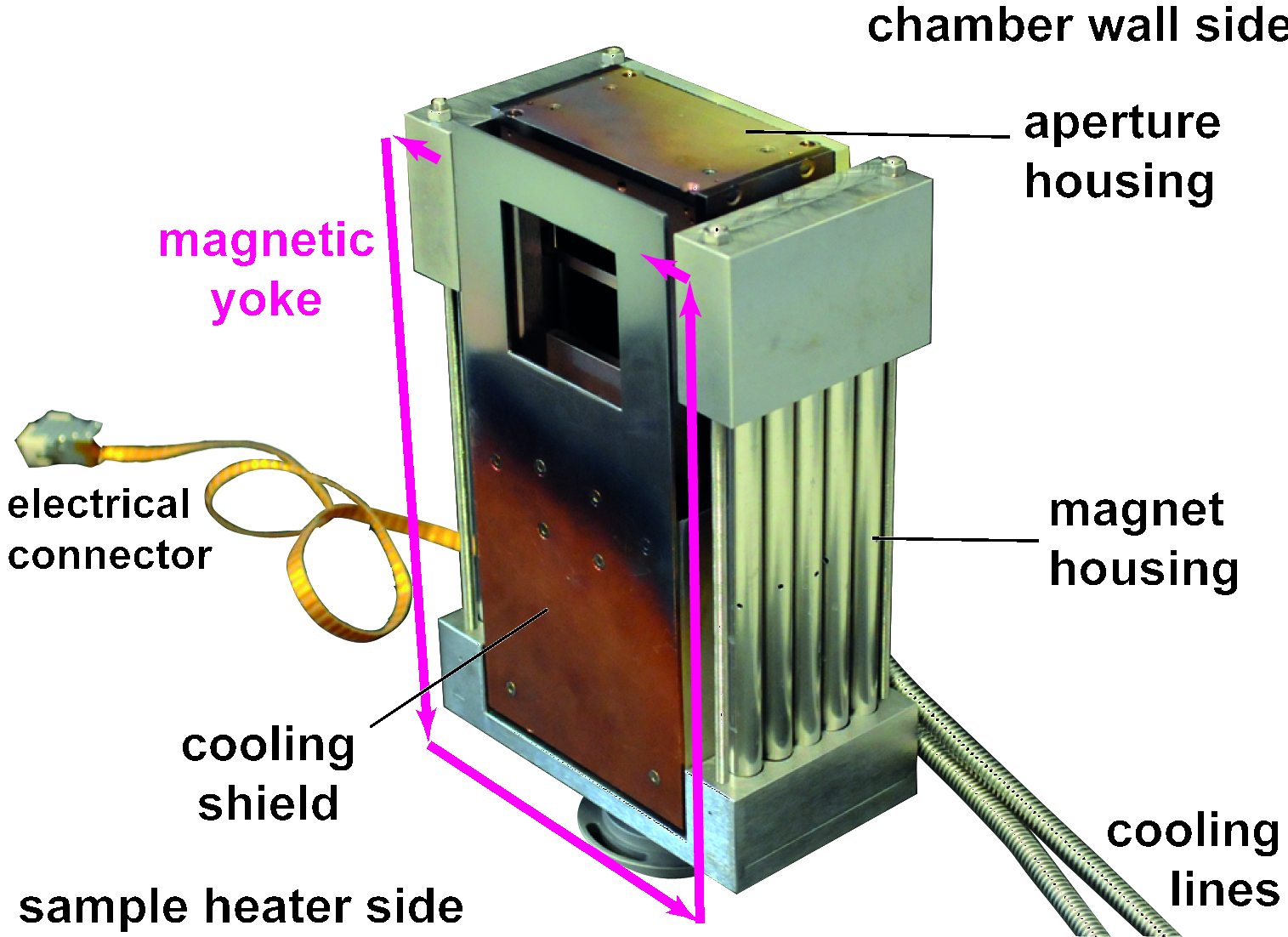}
	\caption{Photograph of the aperture yoke with installed \textit{in situ} neutron aperture. Both aperture and yoke are mounted to a water-cooled copper base. The additional front shield provides additional protection of the aperture housing and mechanics.}
	\label{Fig13}
\end{figure}

To suppress potential diffuse neutron background, which would reduce the signal-to-noise ratio, and with the intention of ideally defining the beam size, a neutron aperture was installed as part of the yoke arrangement directly before the sample heater (Fig.\ \ref{Fig13}), yet still in the guide field path. The aperture consists of two horizontal 3\,mm thick B$_4$C blades with precision ground knife edges. The blades can independently be positioned along the z-axis using two vacuum compatible linear servo drives (50\,mm travel range), from fully closed (slightly overlapping knife edges) to fully open ($60 \times 50$\,mm$^2$ window). 
The linear drives and the blade mechanics are protected from the sputtering environment using a stainless-steel housing that is attached to a water-cooled copper base. It also acts as the heat sink for the guide field yoke surrounding the aperture to prevent the thermally induced demagnetization of the NdFeB magnets above 80$^{\circ}$\,C. An additional copper shield, again thermally connected to the water-cooled copper base, further reduces the heat load on the setup originating from the sample heater and the sputter plasma. The whole yoke-aperture arrangement is mounted to the chamber base.

\subsubsection{Alignment field - Helmhotz coil arrangement}
\begin{figure}
	\includegraphics[width= 0.9 \columnwidth]{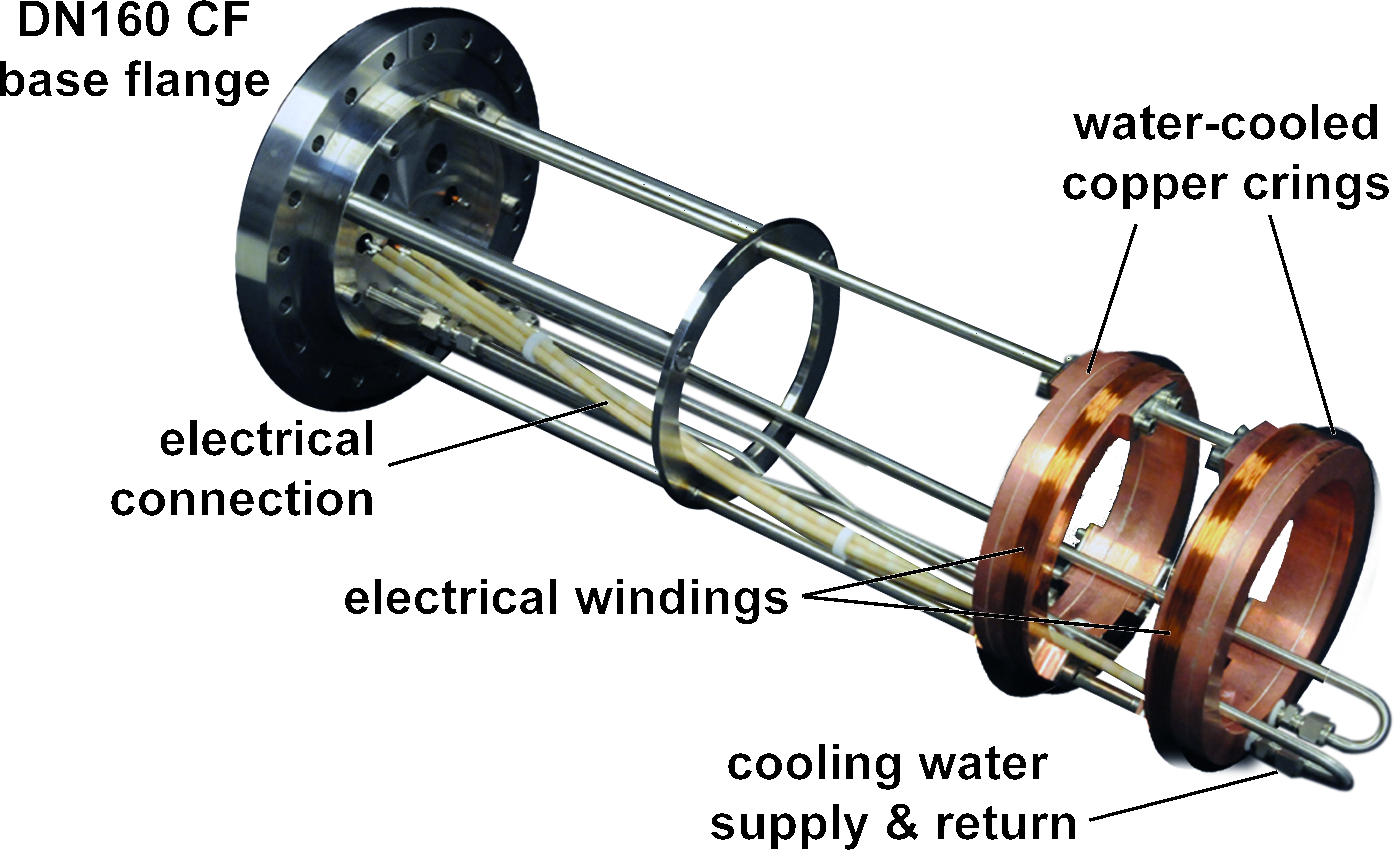}
	\caption{Photograph of the Helmholtz coil. The two copper rings are internally water-cooled and support the electrical winding. The wire winding shown in this photo was later replaced by copper tape windings which have a better packing factor and thermal coupling to the heat sink. The whole setup is mounted on a DN160CF base flange.}
	\label{Fig14}
\end{figure}

\begin{figure}
	\includegraphics[width=\columnwidth]{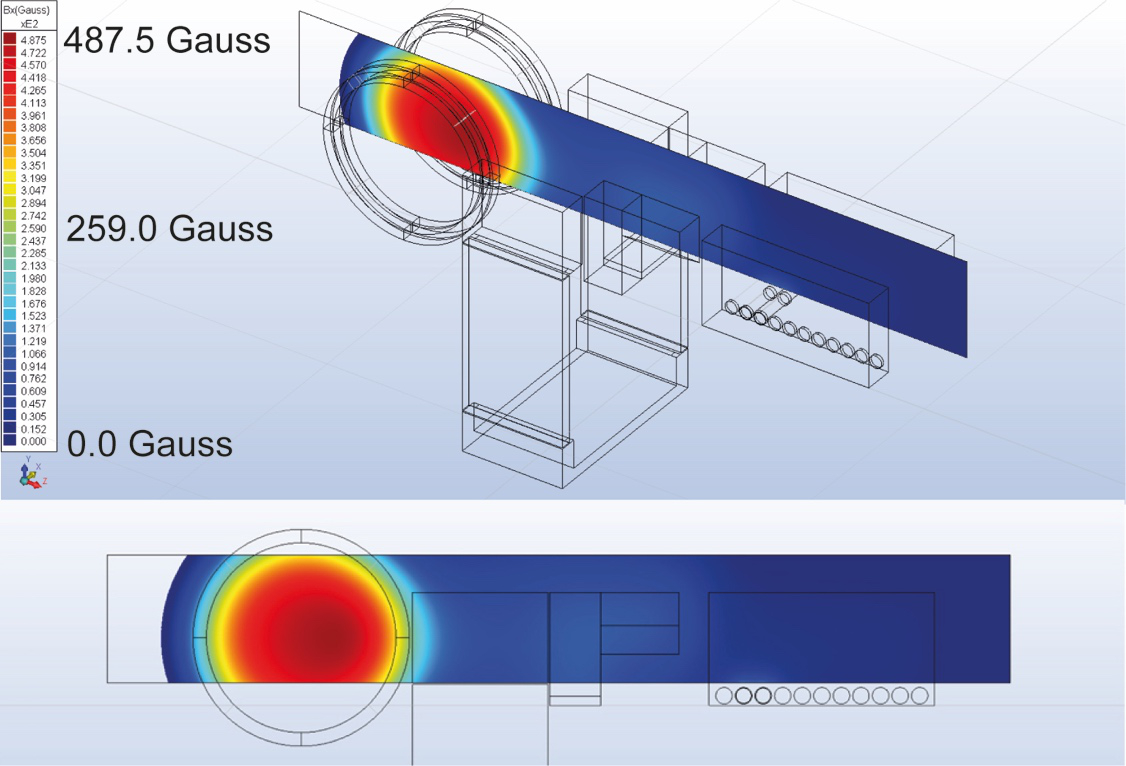}
	\caption{FE simulation of the magnetic field distribution along the guide field yokes and in the Helmholtz coil. A very homogenous constant field strength of about 10\,G in between the yokes is achieved. The maximum design field in the Helmholtz coil reaches almost 500\,G.}
	\label{Fig15}
\end{figure}

\begin{figure}
	\includegraphics[width=\columnwidth]{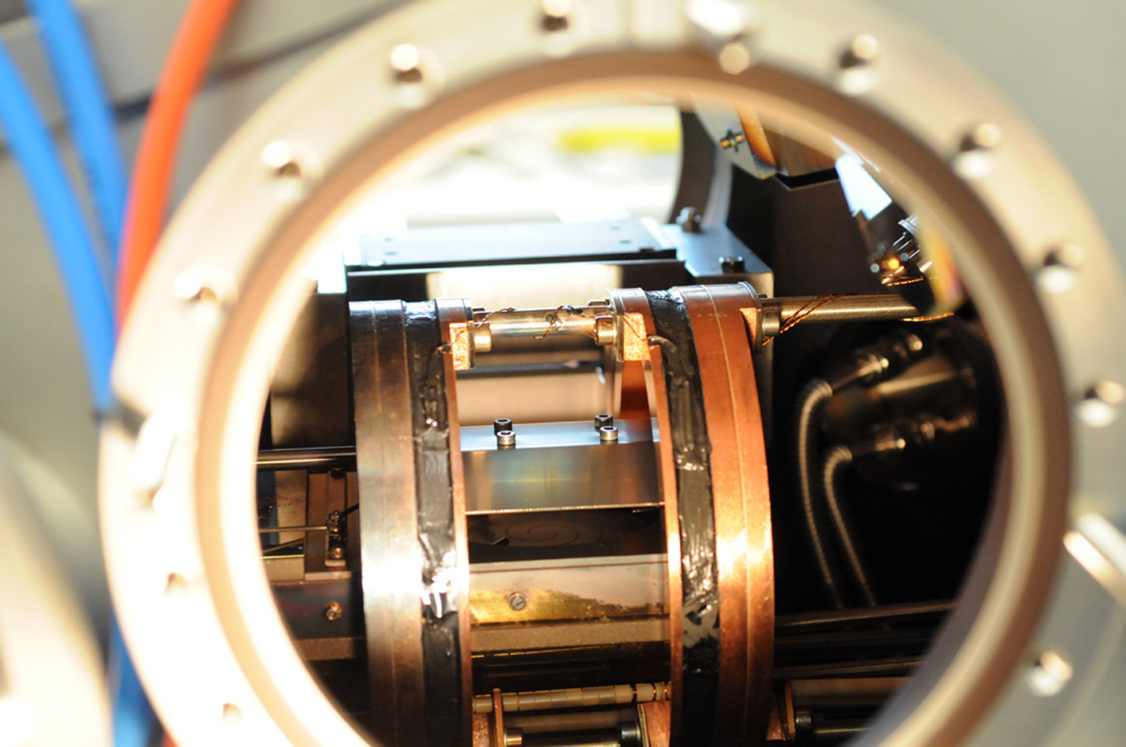}
	\caption{Photograph of the installed magnetic guidance and Helmholtz setup. The black regions on the Helmholtz coil cooling rings originate from the coil package being embedded into Stycast resin for optimized electrical insulation and thermal contact.}
	\label{Fig16}
\end{figure}

\begin{figure}
	\includegraphics[width=\columnwidth]{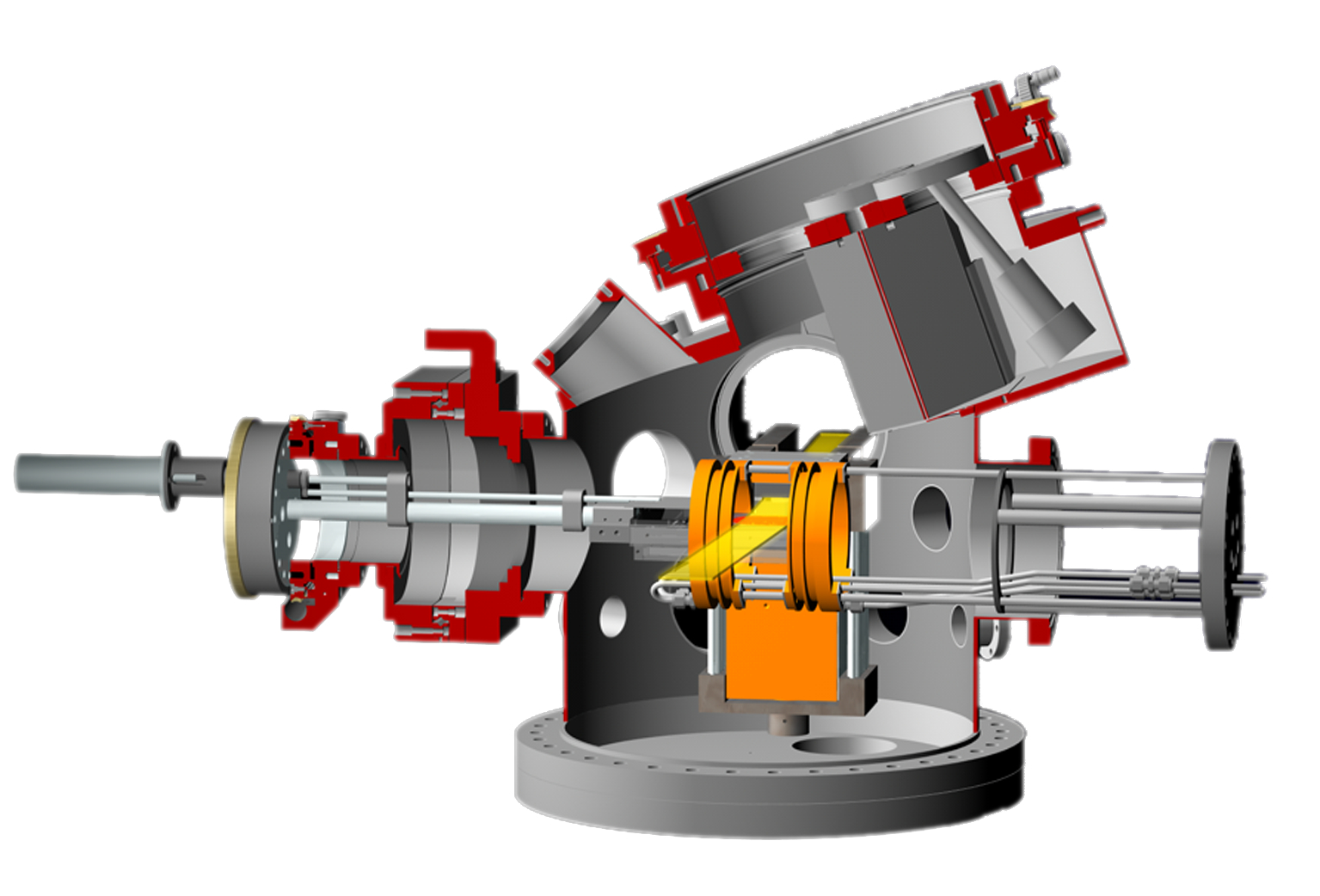}
	\caption{Cut CAD rendering of the deposition system with magnetic guidance and Helmholtz setup. The Helmholtz coil is mounted on a linear vacuum manipulator (not shown), that allows to retract the coil from the sample position for the coating processes.}
	\label{Fig17}
\end{figure}

As PNR is a very sensitive probe for magnetic structures, it is highly desirable to align the moments of sputtered magnetic films.
For typically soft-magnetic films, field strengths of up to 500\,G are usually sufficient. To supply defined magnetic fields for the PNR measurements, we designed a Helmholtz coil arrangement that surrounds the sample heater and does not block the neutron beam. 
The design is based on ring shaped copper coil supports with inner water cooling (Fig.\ \ref{Fig14}). To realize this cooling, trenches were turned into the rings which were then sealed with copper lids in a standard vacuum soldering process. The copper rings have additional trenches in which the electrical copper windings for the field generation are wound. The copper windings are made from single-sided Kapton laminated copper tapes with a cross-section of $10 \times 0.25$\,mm$^2$. This maximizes the fill factor and realizes the best possible thermal contact of the winding package to the cooling ring. For additional electrical insulation and improved thermal contact, the package was finally embedded in UHV-compatible Stycast\,2850 two-component resin \cite{Stycast2850} using a vacuum infusion technique. The cooling rings with the coil packages are supported by three stainless-steel struts. Cooling water is supplied using 6\,mm-diameter stainless-steel tubes. Electrical power is supplied using copper wires with ceramic insulators. The Helmholtz coil arrangement is mounted on a DN160CF vacuum base flange (Fig.\ \ref{Fig14}). To extend the coil into the chamber as well as to retract it again, the coil is mounted on a linear vacuum manipulator with 300\,mm travel range. The latter is attached to the 160\,mm-diameter vacuum port facing the sample heater port. As the linear manipulator extends too far out of the footprint allowed for the installation of the deposition system through the top hatch of the REFSANS experiment, the Helmholtz coil and its manipulator are the only module that have to be installed on the deposition system after it has been placed in the experiment area.     
Like for the guide fields, the magnetic design of the Helmholtz coil was obtained using finite element (FE) Maxwell equation simulations (Fig.\ \ref{Fig15}) which were verified using Hall sensor measurements. The reachable magnetic field strengths are limited by the maximal operation temperature of the Stycast resin of about 130$^{\circ}$\,C (Catalyst 9 \cite{Stycast2850}). When in vacuum, this temperature is reached at about 2.4\,kW input power which corresponds to $\sim 450$\,G field strength, which is close to the desired value of 500\,G. To avoid thermal breakdown of the Helmholtz coil, a temperature monitoring system based on PT100 resistors and interlocks have been implemented. The complete magnetic setup (guidance yokes and Helmholtz coil) is shown in Fig.\ \ref{Fig16} \& \ref{Fig17}.

\subsection{System Mobility}   
\begin{figure}
	\includegraphics[width=\columnwidth]{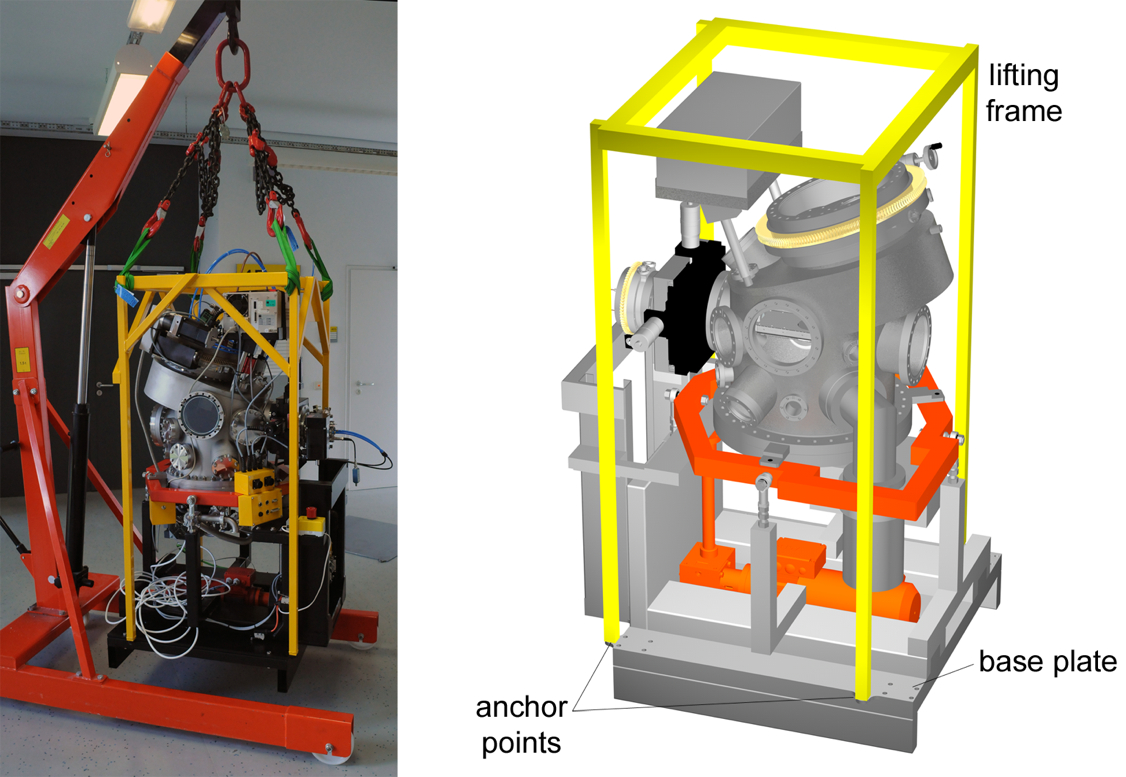}
	\caption{Photograph (left) and CAD rendering (right) of the deposition system with its lifting frame. The frame allows the safe positioning of the system using the gantry crane of the beamline hall.}
	\label{Fig18}
\end{figure}

The deposition system with a total mass of about 650\,kg has to be transported by a gantry crane. Because one can easily pull the system out of the guide rails of the modified fork lift assembly, simple hoisting leashes attached to the upper parts of the system cannot be employed. Therefore, the attachment points of the hoisting gear are located at the outer edges of the base plate of the deposition system. Most of the system’s weight and therefore it's center of gravity is concentrated in the upper third of the setup. If flexible hoisting leashes were used, small lateral forces would therefore be enough to lead to an overturning of the complete system, with the respective hazards for the operators and the equipment. To rule out this risk, a steel frame has been fabricated that is tightly screwed to the base plate of the system (Fig.\ \ref{Fig18}). The frame is welded from $3 \times 3$\,cm$^2$ square bar steel with a wall strength of 1\,mm. Connections to the base plate are provided by four M8 high-performance steel screws with corresponding high-performance steel nuts. To prevent corrosion, the steel frame is powder coated. Through diagonal struts, the lifting frame is rigid enough to accept even large lateral forces and thereby prevents an overturning of the system. The crane is connected to the top of this lift frame using hoisting leashes and lifting chains\footnote{To comply with German safety regulations for hoisting gear and to ensure that the custom made lifting frame can accept the large forces during the lifting, its safety was certified by the German technical surveillance service (Technischer \"Uberwachungsverein, T\"UV).}. 

\subsection{Auxiliary equipment}
A set of auxiliary support equipment completes the deposition system. As these components are also installed near the neutron reflectometer, they are subject to the same mobility, neutron protection and size requirements as the main system. In the following, they are introduced briefly. 

\subsubsection{Gas Handling}
\begin{figure}
	\includegraphics[width=\columnwidth]{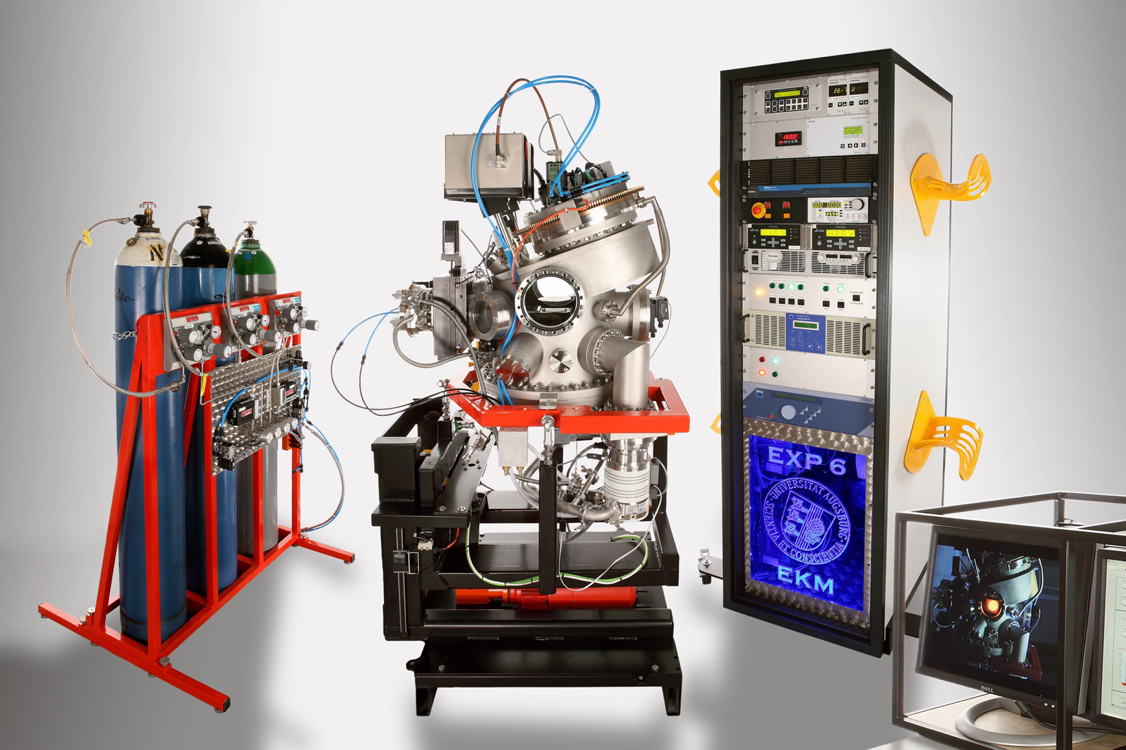}
	\caption{Photograph of the complete deposition system with all auxiliary components. The main setup (middle) includes the vacuum system, and the lift assembly (without Helmholtz coil manipulator). Technical gases are supplied by the three standard bottles which are mounted in a support frame (left). Latter also hosts the gas handling module. The instrumentation and power supplies are located in a 19\,\inch\ rack (right), that is equipped with reels which can accept all cables that connect the system with its electronics. The system is remotely controlled via a computer module (front right corner), which is installed in a steel frame to allow for easy transportation.}
	\label{Fig19}
\end{figure}

For inert sputtering, controlled argon working gas atmospheres have to be established within the deposition system.  For reactive sputtering mixtures of argon/oxygen or argon/nitrogen are used. Furthermore, the system has to be vented with oxygen for oxygen anneal processes or with nitrogen, when the vacuum has to be broken. The three gases (argon, nitrogen, oxygen) are supplied in standard 50\,l, 200\,bar gas bottles. For these, a separate frame was realized, which can be placed on the hallway next to the sputter system at the REFSANS experiment area. It is designed to accept three standard bottles and protects them from falling over (Fig.\ \ref{Fig19}, left). 
The gas atmosphere in the vacuum system is controlled by setting constant flow rates of the required gasses and by controlling the pumping speed of the turbo pump using a motorized controlled gate valve which is automatically adjusted by a pressure controller. Both, turbo pump and gate valve, are part of the main setup. The flow rates of the gases are adjusted using two commercial remote controllable massflow controllers\footnote{MKSmass-flowcontrollertype: 1179A21CM1AV}. To adjust gas mixtures and to completely shut off the gas flows, remotely, four pneumatically actuated needle valves are implemented for the individual gas lines (low line 1, flow line 2, a nitrogen venting, and an oxygen venting line). The needle valves with their respective solenoid switches, the two flow controllers and the required pressure regulators for the three gas bottles were combined into one gas handling module. It is mounted on a steel plate attached to the gas bottle frame. All gas lines are funneled together into one main gas line, that is connected to the vacuum system via a flexible, high-purity gas tube and all-metal, vacuum-tight, and oxygen-proof quick connectors\footnote{Swageloktype: SS-QC4-B1-6MO-SC11}. The gas handling module also hosts the pressure reducer for the compressed air feed of the system.

\subsubsection{Instrumentation Rack}
For the operation and the control of the sputter deposition system, a multitude of additional electronics is required, which is installed in a standard 19\,\inch\ instrumentation rack (Fig.\ \ref{Fig19}, right). The electronics comprises two pressure gauge displays, two flow controllers, a pressure controller, the turbo pump drive unit, a temperature display, power supplies for the heater, RF, and DC sputtering, the servo motor, Helmholtz coil, and stepper motor drives with their power supplies, the controls for the pneumatic valves of the gas handling module, the bypass and the sample shutter controls. The distance between rack and sputter deposition system of about 10\,m is bridged with over 20 custom fabricated cables which connect the individual control and display units with the respective equipment. For the transport of the rack, the total of about 200\,m cable is stored on two reels, attached to the left and right sides of the rack. For fast set up and disassembly, special connector boxes on the deposition system were installed that host the plugs for the control cables. With two exceptions, all cables end in different plug types, minimizing the risk of wrong connections. For ground transportation, the rack is equipped with four wheels. As ground level access to the set-up position of the rack might not be possible, the rack is also movable using a crane. For that purpose, four fixtures near the rack base have been installed to which lift leashes are attached. These leashes are wound around a rectangular steel frame that is put around the top part of the rack to prevent its overturning when lifted. 

\subsubsection{Computer Module}
As the deposition system is not readily accessible during the neutron reflectometry experiments, it can be remote controlled via a computer. A standard PC with USB and RS-232 buses was chosen. It is complemented with a PCI relays card\footnote{Kolter Electronics PCI-OptoRel-Spezial, $16 \times \text{Opto-In}$, $8 \times 2$\,A relays} that comprises eight 2\,A relays and 16 optically coupled inputs for status and interlock read outs. To allow for easy transport and installation, the computer, its two 17\,\inch\ displays, a USB backup hard disk and all power supplies are installed in a rigid steel frame (Fig.\ \ref{Fig19}, front right corner). This computer module can quickly be connected and disconnected from the instrumentation rack and can easily be moved via the experiment halls gantry crane.

\subsubsection{Sample Preparation Module}
To minimize distances and to speed up the transport and insertion of a freshly prepared substrate into the vacuum chamber, the respective equipment for handling, cleaning and sample preparation was combined into a dedicated sample preparation module. Like the computer module, it is based on a rigid steel frame housing and can be placed directly at the beamline and in the vicinity of the deposition system. It contains a hotplate, an ultrasonic cleaner, compartments for organic solvents, tools, beakers, sample and substrate storage, and a connection for a nitrogen blowgun. 

\subsubsection{Water Cooling}
To prevent overheating and provide temperature stability during operation, the sputter sources, the RF power supply, the quartz crystal monitor, the Helmholtz coil, and the turbo pump\footnote{In standby mode, when the turbo pump is only on to maintain the base pressure of the system, air cooling via fans is sufficient. Only under gas load during the sputter deposition process, water cooling of the pump is required.} are water-cooled. The cooling water is provided externally from the facility cooling water circuit. The deposition system as well as the instrumentation rack are connected to the cooling water using quick-connectors. From manifolds, the water is then distributed to the various components.

\subsection{Remote Control and Programming}
The deposition system can be remote controlled via the computer module. Because of size limitations, not all movable parts within the vacuum system (sample shutter, heater, Helmholtz coil and sources) could be designed in a way that inherently rules out any collisions between them. Accidental irradiation of chamber parts with neutrons and an exceeding of the safe travel distance of the system’s lift drive represent potential accident hazards. To minimize the risk for the experimenters and the hardware, great care has been put into the development of the control software of the system. All control programs are based on the programming language LabView. 

\subsubsection{Stepper Motor Control}
All movable interior parts of the deposition system are actuated via stepper motors. Therefore, their control has been combined into one single program. Most of the various axes (i.e.\ sample rotation, x-y-translation, and source rotation) are, however, not equipped with absolute position encoders. For these axes, positioning is only possible through relative drives with respect to known reference points. The individual reference points have to be defined in an initialization procedure that precedes the start of the program's main sequence. After the initialization and after all consequent motions, the axes positions are stored in a text file. This way, even if the control computer needs to be re-started, an initialization only becomes necessary if an axis position has been changed manually during the shut-down.

The heater rotation (d-axis) allows for two pre-defined positions. In the sample position, the heater side faces the sources for the deposition process. In the QCM position, the heater assembly is rotated by 180$^{\circ}$ and the quartz crystal sensor faces the sources for the rate measurements. In the heater position it is allowed to incline the sample by an additional $\pm 20$$^{\circ}$ to adjust the incidence angle of the neutron beam. For accuracy reasons, the sample rotation (d-axis) was equipped with an optical absolute encoder.

\subsubsection{Lift Drive Control}
The servo motor lift drive for the height adjustment of the system is connected via a USB to CAN bus and controlled via an independent program. The lift drive only allows for relative motions with respect to a reference zero position, acquired by an automatic referencing sequence. The referencing is an integrated option of the servo motor controller and can be initiated via the control program. The servo motor controller accepts absolute position values and the reference is preserved even if the control software is quit, provided the servo motor controller is not disconnected from its power supply.

The absolute position of the system is measured by an RS-232 controlled electronic caliper. After the automatic referencing procedure, when the servo motor drive has reached its zero position, the caliper position is read. This height value is stored as offset position, which is then subtracted from the actual position. The corrected value is displayed by the control software. Thereby the zero position of the servo drive and the caliper coincide.

\subsubsection{Substrate Heater Control}
The heater temperature is measured optically using an infrared pyrometer with a lower limit of 160$^{\circ}$\,C. The pyrometer points at the sample puck through a viewport that is located at the blank source position of the DN300CF rotational feedthrough. To avoid accidental overheating and a consequent damaging of the heater, the control software can only be initiated when the pyrometer and the shutter are in the correct positions. This is checked via the position text files of the stepper motors and of the sample shutter. The heater temperature control is based on a standard PID algorithm, that has been modified to account for the specific requirements of the equipment employed. Details on its working principle are given in the supplementary material. For the monitoring of the control parameters, the time evolution of the temperature set point, the measured temperature and the output current are displayed in real time on the control computer. In addition, the heater control software can follow pre-programmed, complex temperature curves, e.g.\ for annealing and heat treatment sequences. The temperature reading was calibrated using a thermocouple for the bare heater surface without any sample placed onto it. Due to different surface emissivities of different samples or surface coatings, material dependent emissivity curve must be taken into account if precise temperature readings are required.

\subsubsection{Main Control Program (MCP)}
The operation of a sputter deposition system requires the simultaneous monitoring and active manipulation of a multitude of essential system parameters. This task has been combined into one main control program (MCP). The various control functions are programmed as independent sub modules, so called virtual instruments (VIs), which are embedded into the main routine of the MCP.

All electronic components that are controlled via the MCP are connected to the control computer via a single USB to $16 \times \text{RS-232}$ hub in the 19\,\inch\ instrumentation rack or are triggered via the $8 \times \text{PCI relays card}$ (gas control, shutter, and bypass). This minimizes the number of bus systems and cabling needed for the physical connection between the computer and the electronics. On the other hand, as the $\text{RS-232}$ hub scans through the individual ports, the  bus system is intrinsically slow. A complete port scan takes about 0.2\,s, which represents the average minimum access time for a request to a certain port (read or write) to which the communication times between the computer and the equipment are added. The latter depend on the baud rates of the addressed electronics. The maximum total time for a complete iteration of all requests to all components is about $1 - 1.5$\,s. This represents the time between readouts of system parameters and execution of a command by the system. As these functions are not time critical, this disadvantage of the chosen bus solution is acceptable.

To nevertheless minimize bus traffic and thus speed up the communication of the control program, only the basic functions like turbo pump and pressure control are active at all times. Other modules like the controls for the rate monitor, the sputtering power supplies, the mass flow controllers, or the pressure controller are only loaded when needed. To further minimize bus traffic, program inputs are dealt with via an event handling architecture which only generates outputs to the bus when the respective program input values are changed. In addition, the internal structure of the MCP was programmed to support parallel polling. Here, the individual request and inputs are handled in parallel, and are dealt with when the respective COM port is available. This enhances the responsiveness of the MCP substantially.

The pneumatic needle valves of the gas handling module, the pneumatic bypass valve and the sample shutter of the deposition system can either be controlled via the MCP or set by physical switches on the 19\,\inch\ instrumentation rack. To avoid contradicting inputs, a selector switch has been implemented to choose between manual and remote control of the switches. In the current version, the MCP only acts as a virtual representation of these physical control switches on the 19\,\inch\ instrumentation rack. Nevertheless, the modular programming of the MCP allows for the easy expansion of the program to run pre-programmed control sequences for highly automated growth processes. The implementation of these functionalities will follow according to arising requirements.

\subsection{Safety}
The safe operation of the deposition system in a beamline as well as its stand-alone use demands a variety of precautions that arise from the use of high temperatures in oxidizing atmospheres, sputtering with high- power/high-voltage power supplies, the vacuum environment and the remote-controlled manipulation of the in-vacuum components, as well as the deposition system itself. Furthermore, the implementation of the system in a neutron beam line with the accompanying radiation hazards demands special provisions that minimize the risk for the operators and the equipment, which are briefly described in the following.

\subsubsection{Emergency Switches}
The main hazards in the operation of the deposition system are electrical shocks due to malfunctions of the high voltage power supplies, as well as mechanical hazards to the operators and the system during the positioning of the vacuum chamber by the lift drive. For the direct shut down of these functions, the system is equipped with three independent emergency-stop switches. These toggle circuit breakers in the 19\,\inch\ instrumentation rack to cut off the power to the high voltage power supplies (DC- and RF sputtering, heater) and the motor drive. To ensure accessibility of the switches from all possible operation positions, one switch is located at the computer, one on the instrumentation rack and one on the mounting frame of the vacuum chamber. Vital system functions like the vacuum pump control and gas handling are not affected by emergency shut offs.

Due to the radiation hazards, a direct observation of the system by an operator during reflectometry measurements is not possible. To avoid mechanical damage to the system by exceeding the height limit of the lift drive, two microswitches are toggled by the moving part of the vacuum chamber mounting frame. The first switch acts as conventional end position switch that sends a stop signal to the frequency convertor of the servo motor lift drive. In case of a failure of this first switch, a second one has been installed slightly above, that acts as emergency shut off switch, acting on the circuit breaker in the instrumentation rack. 
Other functions of the system (sources, heater) are not affected by these measures.

\subsubsection{Position Control}
The motion of the DN100CF and the DN300CF rotational feedthroughs is restricted by the electrical and cooling water connections to the heater and the sputter sources. To avoid twisting of these connections, the angular freedom was limited to 270$^{\circ}$. These limits are implemented in the control software of the respective stepper motor drives. To be effective, the respective rotation axis has to be referenced correctly.

The translation of the heater along the x- and z-axis, the source rotation, and the shutter operation can potentially lead to collisions of these in-vacuum components of the system. To avoid this, the respective control software coordinates the individual motions of the equipment and ensures collision free operation, provided the stepper motor axes have been referenced correctly.

In addition to the positioning safeguards implemented in the control software, visual supervision of the interior of the vacuum chamber is provided by a webcam through a vacuum viewport on the top of the system. The real-time pictures are displayed on the control computer.

\subsubsection{Radiation Safety}
Even though grate care has been taken to provide passive safety from radioactive activation of system components by neutron shields, there do exist source and heater positions that would potentially allow a contamination of the sources or the rate monitor sensor to occur. To prevent this, the system is equipped with an interlock switch, that can be connected to the main neutron shutter of the beamline. Only when this interlock is closed, the neutron beam can be switched on. The interlock can be operated manually by a switch on the instrumentation rack or it can be remotely controlled via the main control program for which, one of the PCI card relay outputs is assigned to act as interlock for the neutron beam line shutter. The latter option also offers the possibility to only open the interlock, when allowed by the MCP, minimizing the risk of human error. A computer controlled shut off of the beam or the system, that is triggered by this second interlock is also implemented as part of the MCP.

\section{Summary and Conclusions}
We have built a versatile sputter deposition system for the \textit{in situ}- and \textit{in-operando} use in PNR experiments. The system allows the safe and time-efficient deposition of films and heterostructures containing up to three source materials. The three sputter gasses allow the use of a multitude of material classes, encompassing pure elements, oxides, and nitrides. The samples can stay aligned in the measurement position at all times allowing for quick alternations between deposition and measurement. This helps in avoiding potential measurement errors originating form contamination, degradation, or misalignment.

The system is designed fully mobile with a total footprint of the deposition setup of less than $1 \times 1$\,m$^2$. This allows for the installation of the system in beamlines that are not permanently dedicated for \textit{in situ} experiments. With installation times of $\sim 2$\,h, the setup is quickly utilizable for experiments, saving costly experiment time.

Originally designed for the REFSANS experiment at the MLZ in Garching, the compact and flexible design allows the use of the system in virtually any beamline with suitable geometry.  Originally commissioned, tested and optimized at REFSANS, the first published experimental data using the deposition system was in fact obtained at the AMOR beamline at Paul Scherrer Institute, Villigen, Switzerland \cite{doi:10.1103/PhysRevApplied.7.054004}.



\section*{Acknowledgements}
We acknowledge the very helpful discussions with Jean-Fran\c{c}ois Moulin and Martin H\"ase of the REFSANS team, as well as their support during the commissioning experiments of the deposition system. The work was supported by the Deutsche Forschungsgemeinschaft via the Transregional Research Center TRR\,80.
J.\ Mannhart acknowledges support by the DFG (Leibniz prize grant).

\section*{References}
\bibliography{A_Schmehl}

\begin{thebibliography}{10}
\expandafter\ifx\csname url\endcsname\relax
  \def\url#1{\texttt{#1}}\fi
\expandafter\ifx\csname urlprefix\endcsname\relax\def\urlprefix{URL }\fi
\expandafter\ifx\csname href\endcsname\relax
  \def\href#1#2{#2} \def\path#1{#1}\fi

\bibitem{10.1038/ncomms5530}
Y.~F. Nie, Y.~Zhu, C.-H. Lee, L.~F. Kourkoutis, J.~A. Mundy, J.~Junquera,
  P.~Ghosez, D.~J. Baek, S.~Sung, X.~X. Xi, K.~M. Shen, D.~A. Muller, D.~G.
  Schlom, Atomically precise interfaces from non-stoichiometric deposition,
  Nat. Commun. 5 (2014) 4530.
\newblock \href {http://dx.doi.org/10.1038/ncomms5530}
  {\path{doi:10.1038/ncomms5530}}.

\bibitem{doi:10.1146/annurev-matsci-070115-032057}
S.~Middey, J.~Chakhalian, P.~Mahadevan, J.~Freeland, A.~Millis, D.~Sarma,
  Physics of ultrathin films and heterostructures of rare-earth nickelates,
  Annual Review of Materials Research 46~(1) (2016) 305--334.
\newblock \href {http://dx.doi.org/10.1146/annurev-matsci-070115-032057}
  {\path{doi:10.1146/annurev-matsci-070115-032057}}.

\bibitem{doi:10.1021/acs.nanolett.5b02720}
M.~Gibert, M.~Viret, A.~Torres-Pardo, C.~Piamonteze, P.~Zubko, N.~Jaouen, J.-M.
  Tonnerre, A.~Mougin, J.~Fowlie, S.~Catalano, A.~Gloter, O.~St\'ephan, J.-M.
  Triscone, Interfacial control of magnetic properties at lamno3/lanio3
  interfaces, Nano Letters 15~(11) (2015) 7355--7361.
\newblock \href {http://dx.doi.org/10.1021/acs.nanolett.5b02720}
  {\path{doi:10.1021/acs.nanolett.5b02720}}.

\bibitem{doi:10.1143/JJAP.29.L1199}
M.~Yoshimoto, H.~Nagata, T.~Tsukahara, H.~Koinuma,
  \href{http://stacks.iop.org/1347-4065/29/i=7A/a=L1199}{In situ rheed
  observation of ceo 2 film growth on si by laser ablation deposition in
  ultrahigh-vacuum}, Japanese Journal of Applied Physics 29~(7A) (1990) L1199.
\newline\urlprefix\url{http://stacks.iop.org/1347-4065/29/i=7A/a=L1199}

\bibitem{doi:10.1063/1.1384432}
W.~Matz, N.~Schell, W.~Neumann, J.~B{\o}ttiger, J.~Chevallier, A two magnetron
  sputter deposition chamber for in situ observation of thin film growth by
  synchrotron radiation scattering, Review of Scientific Instruments 72~(8)
  (2001) 3344--3348.
\newblock \href {http://dx.doi.org/10.1063/1.1384432}
  {\path{doi:10.1063/1.1384432}}.

\bibitem{doi:10.1103/PhysRevLett.64.2929}
O.~M. Magnussen, J.~Hotlos, R.~J. Nichols, D.~M. Kolb, R.~J. Behm,
  \href{https://link.aps.org/doi/10.1103/PhysRevLett.64.2929}{Atomic structure
  of cu adlayers on au(100) and au(111) electrodes observed by in situ scanning
  tunneling microscopy}, Phys. Rev. Lett. 64 (1990) 2929--2932.
\newblock \href {http://dx.doi.org/10.1103/PhysRevLett.64.2929}
  {\path{doi:10.1103/PhysRevLett.64.2929}}.
\newline\urlprefix\url{https://link.aps.org/doi/10.1103/PhysRevLett.64.2929}

\bibitem{doi:10.1021/ma0119833}
P.~Lavalle, C.~Gergely, F.~J.~G. Cuisinier, G.~Decher, P.~Schaaf, J.~C. Voegel,
  C.~Picart, Comparison of the structure of polyelectrolyte multilayer films
  exhibiting a linear and an exponential growth regime: An in situ atomic force
  microscopy study, Macromolecules 35~(11) (2002) 4458--4465.
\newblock \href {http://dx.doi.org/10.1021/ma0119833}
  {\path{doi:10.1021/ma0119833}}.

\bibitem{doi:10.1063/1.4972993}
A.~S. Mohd, S.~P\"utter, S.~Mattauch, A.~Koutsioubas, H.~Schneider, A.~Weber,
  T.~Br\"uckel, A versatile uhv transport and measurement chamber for neutron
  reflectometry under uhv conditions, Review of Scientific Instruments 87~(12)
  (2016) 123909.
\newblock \href {http://dx.doi.org/10.1063/1.4972993}
  {\path{doi:10.1063/1.4972993}}.

\bibitem{doi:10.1063/1.3169506}
J.~A. Dura, J.~LaRock, A molecular beam epitaxy facility for in situ neutron
  scattering, Review of Scientific Instruments 80~(7) (2009) 073906.
\newblock \href {http://dx.doi.org/10.1063/1.3169506}
  {\path{doi:10.1063/1.3169506}}.

\bibitem{doi:10.1051/epjap/2012110295}
J.~Stahn, U.~Filges, T.~Panzner,
  \href{https://doi.org/10.1051/epjap/2012110295}{Focusing specular neutron
  reflectometry for small samples}, Eur. Phys. J. Appl. Phys. 58~(1) (2012)
  11001.
\newblock \href {http://dx.doi.org/10.1051/epjap/2012110295}
  {\path{doi:10.1051/epjap/2012110295}}.
\newline\urlprefix\url{https://doi.org/10.1051/epjap/2012110295}

\bibitem{doi:10.1016/j.nima.2016.03.007}
J.~Stahn, A.~Glavic,
  \href{http://www.sciencedirect.com/science/article/pii/S0168900216300250}{Focusing
  neutron reflectometry: Implementation and experience on the tof-reflectometer
  amor}, Nuclear Instruments and Methods in Physics Research Section A:
  Accelerators, Spectrometers, Detectors and Associated Equipment 821 (2016) 44
  -- 54.
\newblock \href {http://dx.doi.org/10.1016/j.nima.2016.03.007}
  {\path{doi:10.1016/j.nima.2016.03.007}}.
\newline\urlprefix\url{http://www.sciencedirect.com/science/article/pii/S0168900216300250}

\bibitem{doi:10.1103/PhysRevApplied.7.054004}
W.~Kreuzpaintner, B.~Wiedemann, J.~Stahn, J.-F. m.~c. Moulin, S.~Mayr,
  T.~Mairoser, A.~Schmehl, A.~Herrnberger, P.~Korelis, M.~Haese, J.~Ye,
  M.~Pomm, P.~B\"oni, J.~Mannhart,
  \href{https://link.aps.org/doi/10.1103/PhysRevApplied.7.054004}{\textit{In
  situ} polarized neutron reflectometry: Epitaxial thin-film growth of fe on
  cu(001) by dc magnetron sputtering}, Phys. Rev. Applied 7 (2017) 054004.
\newblock \href {http://dx.doi.org/10.1103/PhysRevApplied.7.054004}
  {\path{doi:10.1103/PhysRevApplied.7.054004}}.
\newline\urlprefix\url{https://link.aps.org/doi/10.1103/PhysRevApplied.7.054004}

\bibitem{Kampmann20061161}
R.~Kampmann, M.~Haese-Seiller, V.~Kudryashov, B.~Nickel, C.~Daniel, W.~Fenzl,
  A.~Schreyer, E.~Sackmann, J.~R\"adler, {Horizontal ToF-neutron reflectometer
  REFSANS at FRM-II Munich/Germany: First tests and status}, Physica B:
  Condensed Matter 385 - 386~(Part 2) (2006) 1161 -- 1163.
\newblock \href {http://dx.doi.org/10.1016/j.physb.2006.05.399}
  {\path{doi:10.1016/j.physb.2006.05.399}}.

\bibitem{Kampmann2004E763}
R.~Kampmann, M.~Haese-Seiller, V.~Kudryashov, V.~Deriglazov, M.~Tristl,
  C.~Daniel, B.~Toperverg, A.~Schreyer, E.~Sackmann, {The potential of the
  horizontal reflectometer REFSANS/FRM-II for measuring low reflectivity and
  diffuse surface scattering}, Physica B: Condensed Matter 350~(1 -- 3,
  Supplement) (2004) E763 -- E766.
\newblock \href {http://dx.doi.org/10.1016/j.physb.2004.03.198}
  {\path{doi:10.1016/j.physb.2004.03.198}}.

\bibitem{Moulin20151}
J.~F. Moulin, M.~Haese-Seiller, {REFSANS: Reflectometer and evanescent wave
  small angle neutron spectrometer}, Journal of large-scale research facilities
  1~(A9) (2015) 1 -- 3.
\newblock \href {http://dx.doi.org/10.17815/jlsrf-1-31}
  {\path{doi:10.17815/jlsrf-1-31}}.

\bibitem{NeutronDataBooklet}
A.-J. Dianoux, G.~Lander (Eds.),
  \href{https://www.ill.eu/fileadmin/users_files/documents/links/documentation/NeutronDataBooklet.pdf}{Neutron
  Data Booklet}, 2nd Edition, Old City Publishing, Incorporated, 2003.
\newline\urlprefix\url{https://www.ill.eu/fileadmin/users_files/documents/links/documentation/NeutronDataBooklet.pdf}

\bibitem{tectra}
{tectra, Physikalische Instrumente, Reuterweg 65, D-60323 Frankfurt/M,
  Germany}, \href{http://tectra.de/sample-preparation/heaters/}{{tectra,
  Physikalische Instrumente, Reuterweg 65, D-60323 Frankfurt/M, Germany}},
  accessed: 2017-08-06.
\newline\urlprefix\url{http://tectra.de/sample-preparation/heaters/}

\bibitem{Momentiveperformanceproducts}
{Momentive performance products, 22 Corporate Woods Boulevard, Albany, NY
  12211, USA},
  \href{https://momentive.com/en-US/categories/ceramics/heaters-and-heater-assemblies/#}{{Momentive
  performance products, 22 Corporate Woods Boulevard, Albany, NY 12211, USA}},
  accessed: 2017-08-06.
\newline\urlprefix\url{https://momentive.com/en-US/categories/ceramics/heaters-and-heater-assemblies/#}

\bibitem{QCM}
{INFICON Holding AG, Hintergasse 15B, CH-7310 Bad Ragaz, Switzerland},
  \href{http://products.inficon.com/getattachment.axd/?attaname=2017+fl+single+datasheet}{Front
  load single sensor}, accessed: 2017-11-06.
\newline\urlprefix\url{http://products.inficon.com/getattachment.axd/?attaname=2017+fl+single+datasheet}

\bibitem{Stycast2850}
{Emerson and Cuming, Europe, Nijverheidsstraat 7, B-2260 Westerlo, Belgium},
  \href{https://lartpc-docdb.fnal.gov/0000/000059/001/stycas2850.pdf}{Stycast
  2850 resin technical data sheet}, accessed: 2017-08-06.
\newline\urlprefix\url{https://lartpc-docdb.fnal.gov/0000/000059/001/stycas2850.pdf}

\end{thebibliography}

\newpage


\end{document}